\documentclass[a4paper,11pt]{article}
\pdfoutput=1 
\usepackage{jcappub}
\usepackage{amsmath}
\usepackage{graphicx}
\usepackage{latexsym}
\usepackage{xspace}
\usepackage{color}
\usepackage{hyperref} 
\usepackage{bm}
\usepackage{relsize}
\usepackage{tabularx}
\usepackage{multirow}
\usepackage{amssymb}
\usepackage{slashbox}
\usepackage[table]{xcolor}

\makeatletter
\gdef\@fpheader{}
\g@addto@macro\bfseries{\boldmath}
\makeatother

\newcommand{\ie}{{i.e.~}}

\newcommand{\order}[1]{\mathcal{O}\!\left(#1\right)}

\DeclareMathOperator{\Ei}{Ei}

\newcommand{\dd}{\mathrm{d}}
\newcommand{\ee}{e}

\newcommand{\sss}[1]{{\scriptscriptstyle{#1}}}

\newcommand{\uPl}{\mathrm{Pl}}
\newcommand{\uin}{\mathrm{in}}

\newcommand{\umax}{\mathrm{max}}
\newcommand{\uend}{\mathrm{end}}

\newcommand{\ucl}{\mathrm{cl}}

\newcommand{\usssPl}{\sss{\uPl}}

\newcommand{\calP}{\mathcal{P}}

\newcommand{\Mp}{M_\usssPl}

\newcommand{\efolds}{$e$-folds~}

\newcommand{\beq}{\begin{equation}}
\newcommand{\eeq}{\end{equation}}
\newcommand{\bea}{\begin{eqnarray}}
\newcommand{\eea}{\end{eqnarray}}

\newlength{\wsingfig}
\newlength{\wdblefig}
\newlength{\wquadfig}
\newlength{\wtriplefig}
\newcommand{\Eq}[1]{Eq.~(\ref{#1})}
\newcommand{\Eqs}[1]{Eqs.~(\ref{#1})}
\newcommand{\Fig}[1]{Fig.~{\ref{#1}}}

\newcommand{\Ref}[1]{Ref.~{\cite{#1}}}

\newcommand{\Sec}[1]{Sec.~\ref{#1}}
\newcommand{\Secs}[1]{Secs.~\ref{#1}}
\newcommand{\App}[1]{Appendix~\ref{#1}}

\subheader{}

\title{Multiple Fields in Stochastic Inflation}

\author[a]{Hooshyar Assadullahi,}
\author[b]{Hassan Firouzjahi,}
\author[c,b]{Mahdiyar Noorbala,}
\author[a]{Vincent Vennin}
\author[a]{and David Wands}

\affiliation[a]{Institute of Cosmology \& Gravitation, University of Portsmouth, Dennis Sciama Building, Burnaby Road, Portsmouth, PO1 3FX, United Kingdom}
\affiliation[b]{School of Astronomy, Institute for Research in Fundamental Sciences (IPM), Tehran, Iran, P.O. Box 19395-5531}
\affiliation[c]{Department of Physics, University of Tehran, Iran, P.O. Box 14395-547}

\emailAdd{hooshyar.assadullahi@port.ac.uk}
\emailAdd{firouz@ipm.ir}
\emailAdd{mnoorbala@ut.ac.ir}
\emailAdd{vincent.vennin@port.ac.uk}
\emailAdd{david.wands@port.ac.uk}

\date{today}

\begin{document}
\sloppy

\abstract{
Stochastic effects in multi-field inflationary scenarios are investigated. A hierarchy of diffusion equations is derived, the solutions of which yield moments of the numbers of inflationary $e$-folds. Solving the resulting partial differential equations in multi-dimensional field space is more challenging than the single-field case. A few tractable examples are discussed, which show that the number of fields is, in general, a critical parameter. When more than two fields are present for instance, the probability to explore arbitrarily large-field regions of the potential, otherwise inaccessible to single-field dynamics, becomes non-zero. In some configurations, this gives rise to an infinite mean number of $e$-folds, regardless of the initial conditions. Another difference with respect to single-field scenarios is that multi-field stochastic effects can be large even at sub-Planckian energy. This opens interesting new possibilities for probing quantum effects in inflationary dynamics, since the moments of the numbers of $e$-folds can be used to calculate the distribution of primordial density perturbations in the stochastic-$\delta N$ formalism.
}



\maketitle

\section{Introduction}
\label{sec:intro}
Inflation is the leading paradigm to describe the physical conditions that prevailed in the very early Universe~\cite{Starobinsky:1980te, Sato:1980yn, Guth:1980zm, Linde:1981mu, Albrecht:1982wi, Linde:1983gd}. It is a phase of accelerated expansion that solves the puzzles of the standard hot Big Bang model, and provides a causal mechanism for generating scalar~\cite{Mukhanov:1981xt, Hawking:1982cz,  Starobinsky:1982ee, Guth:1982ec, Bardeen:1983qw} and tensor~\cite{Starobinsky:1979ty} inhomogeneous perturbations on cosmological scales. These inhomogeneities result from the parametric amplification of the  vacuum quantum fluctuations of the gravitational and matter fields during the accelerated expansion, that later seed the large scale structure of the Universe. 

The transition from these quantum fluctuations to classical but stochastic density perturbations~\cite{Polarski:1995jg,Lesgourgues:1996jc,Kiefer:2008ku,Martin:2012pea,Burgess:2014eoa,Martin:2015qta} gives rise to the stochastic inflation formalism~\cite{Starobinsky:1982ee, Starobinsky:1986fx, Nambu:1987ef, Nambu:1988je, Kandrup:1988sc, Nakao:1988yi, Nambu:1989uf, Mollerach:1990zf, Linde:1993xx, Starobinsky:1994bd}. It consists of an effective theory for the long-wavelength parts of the quantum fields, which are ``coarse grained'' at a fixed physical scale (\ie non-expanding), larger than the Hubble radius during the whole inflationary period. The non-commutative parts of these coarse grained fields are small (compared to their anti-commutative parts), and at this scale, short-wavelength quantum fluctuations have negligible non-commutative parts too. In this framework, they behave as a classical noise acting on the dynamics of the super-Hubble scales, and the coarse grained fields can thus be described by a stochastic classical theory, following Langevin equations.

The stochastic formalism accounts for the quantum modification of the super-Hubble scales dynamics, and allows us to study how quantum effects modify inflationary observable predictions. Stochastic inflation is indeed a powerful tool for calculating correlation functions of quantum fields during inflation~\cite{GarciaBellido:1993wn, Starobinsky:1994bd, Kahya:2006hc, Finelli:2008zg, Finelli:2010sh, Garbrecht:2013coa, Garbrecht:2014dca, Levasseur:2014ska, Onemli:2015pma, Prokopec:2015owa, Boyanovsky:2015jen, Boyanovsky:2015xoa, Boyanovsky:2015tba, Burgess:2015ajz} (see also \Ref{Prokopec:2007ak, Prokopec:2008gw} for the case of scalar electrodynamics during inflation and \Ref{Tsamis:2005hd} for the case of derivative interactions and constrained fields). In practice, it can be connected to cosmological observations through the $\delta N$ formalism~\cite{Starobinsky:1982ee, Starobinsky:1986fxa, Salopek:1990jq, Sasaki:1995aw, Sasaki:1998ug, Wands:2000dp, Lyth:2004gb, Lyth:2005fi}, which relates the statistical properties of scalar curvature perturbations $\zeta$ to the distribution of the number of \efolds $N$ among a family of homogeneous universes. In the stochastic formalism, this number of \efolds (realised between an initial flat slice of space-time and a final slice of uniform energy density) is a stochastic quantity that we denote $\mathcal{N}$, and its statistical moments directly give rise to correlation functions of cosmological perturbations. For example, the power spectrum of curvature perturbations can be expressed as
\begin{align}
\calP_\zeta=\dfrac{\dd\left(\left\langle \mathcal{N}^2 \right\rangle - \left\langle \mathcal{N} \right\rangle^2\right)}{\dd\left\langle \mathcal{N} \right\rangle}\, ,
\label{eq:Pzeta:Nmean}
\end{align}
where $\mathcal{N}$ is the number of \efolds realised between the time when the scale at which $\calP_\zeta$ is calculated exits the Hubble radius during inflation and the end of inflation, and similar expressions can be derived for higher correlation functions. This is the so-called ``stochastic-$\delta N$ formalism''~\cite{Enqvist:2008kt, Fujita:2013cna, Fujita:2014tja,Vennin:2015hra,Kawasaki:2015ppx}. In fact, this approach may also be called ``stochastic-$N$ formalism'' since it does not rely on an expansion in $\delta N$ and in the metric perturbation $\zeta$ (for instance, these two quantities are not small in the so-called regime of ``eternal inflation'').

The problem therefore boils down to calculating statistical moments of the realised number of \efolds in stochastic inflation. In \Ref{Vennin:2015hra}, this was done for single-field setups where it was shown that in most potentials, stochastic corrections to inflationary predictions remain small within the observational window as long as the inflationary energy scale is sub-Planckian. In this paper, we extend the analysis to multiple field scenarios. Indeed, stochastic dynamics in multiple field potentials can be highly non-trivial~\cite{Clesse:2009ur, Martin:2011ib, Levasseur:2013tja, Fujita:2014tja, Kawasaki:2015ppx}, and has a priori the potential to substantially extend the single-field phenomenology. 

This work designs a generic stochastic-$\delta N$ formalism for multi-field inflation and is organised as follows. In \Sec{sec:FPT:moments}, techniques from ``first passage time analysis'' are employed to calculate the statistical moments of the inflationary numbers of \efolds in multiple field setups. The results obtained in single-field potentials are reviewed, and first indications are given why including more than one scalar field can lead to fundamental differences. In \Sec{sec:harmonic}, ``harmonic potentials'' are introduced, for which the problem can be addressed analytically. In \Sec{sec:InfiniteInflation}, the calculation is carried out for a subclass of harmonic potentials called ``$v(r)$ potentials'', where it is notably shown that, in some cases, including more than one scalar field leads to an infinite mean \efolds number. Implications of this ``infinite inflation'' phenomenon are discussed, in particular studying the probability of probing arbitrarily large-field regions of the potential that are classically impossible to reach otherwise. In \Sec{sec:InhomogeneousEnd}, the case where extra fields modulate the surface of end of inflation is investigated. It is shown that depending on the details of the end-surface, stochastic corrections may be large even at sub-Planckian energy densities, and tend to smear out the effect of the modulating fields. Finally, in \Sec{sec:conclusion}, we present a few concluding remarks, and we end the paper with several appendices containing various technical aspects. 
\section{Moments of Inflationary $e$-folds}
\label{sec:FPT:moments}
We investigate the situation where inflation is driven by $D$ canonical scalar fields $\phi_i$, where $1\leq i\leq D$, slowly rolling down the potential $V(\phi_i)$. Their coarse-grained parts evolve according to the Langevin equations~\cite{Starobinsky:1986fxa}
\begin{align}
\label{eq:Langevin}
\frac{\dd\phi_i}{\dd N}=-\frac{V_{\phi_i}}{3H^2}+\frac{H}{2\pi}\xi_i\, ,
\end{align}
where $\xi_i$ are $D$ independent normalised white Gaussian noises, so that $\langle \xi_i(N_1)\xi_j(N_2) \rangle = \delta_{i,j}\delta(N_1-N_2)$, and $V_{\phi_i}$ denotes the derivative of the potential $V$ with respect to $\phi_i$. Time is labeled by the number of \efolds $N\equiv\ln a$, where $a$ is the scale factor. At leading order in slow roll, the Hubble parameter $H$ is related to the potential energy through the Friedmann equation $H^2=V/(3\Mp^2)$, where $\Mp$ is the reduced Planck mass. In this section, we explain how the statistical moments of the inflationary \efolds generated by \Eq{eq:Langevin} can be calculated. We briefly review the results~\cite{Vennin:2015hra} obtained in single-field setups, and provide hints why including more than one scalar field may a priori lead to fundamental differences.
\subsection{Fokker-Planck Equation}
Let us first recast these stochastic processes through a Fokker-Planck equation, which governs the time evolution of the probability density $P(\phi_i,N)$ that the fields take value $\phi_i$ at time $N$. Introducing the dimensionless potential 
\begin{align}
v=\frac{V}{24\pi^2\Mp^4}\, ,
\end{align}
in the It\^o interpretation\footnote{More generally, the last term in \Eq{eq:FokkerPlanck} can be written in the form
\begin{align}
\sum_i\frac{\partial}{\partial\phi_i}\left[v^\alpha\frac{\partial}{\partial\phi_i}\left(v^{1-\alpha}P\right)\right]
\end{align}
with $0\le \alpha\le 1$, where $\alpha=0$ corresponds to the It\^o interpretation and $\alpha=1/2$ to the Stratonovich one~\cite{Stratonovich:1966}. One can show that keeping terms explicitly dependent on $\alpha$ exceeds the accuracy of the stochastic approach in its leading approximation~(\ref{eq:Langevin}). In particular, corrections to the noise term due to self-interactions of small-scale fluctuations (if they exist) are at least of the same order or even larger. This is why explicit dependence on $\alpha$ is dropped here.}~\cite{Starobinsky:1986fx,Winitzki:1995pg, Vilenkin:1999kd}, it reads
\begin{align}
\frac{1}{\Mp^2}\frac{\partial}{\partial N}P\left(\phi_i,N\right)=\sum_i 
\frac{\partial}{\partial \phi_i}\left[\frac{v_{\phi_i}}{v}P\left(\phi_i,N\right)\right]
+\sum_i\frac{\partial^2}{\partial\phi_i^2}\left[ v P\left(\phi_i,N\right)\right]\, .
\label{eq:FokkerPlanck}
\end{align}
This equation can be written as $\partial P/\partial N \equiv \mathcal{L}_\mathrm{FP}\cdot P \equiv - \bm{\nabla}\cdot \bm{J}$, where $\mathcal{L}_\mathrm{FP}$ is the Fokker-Planck differential operator defined in \Eq{eq:FokkerPlanck}, $\bm{\nabla}$ denotes the vector differential operator $\nabla_i=\partial/\partial\phi_i$ in field space, which, for simplicity, we assume to be flat, and $\bm{J}$ is the probability current, $J_i=\Mp^2[v_{\phi_i}P/v+\partial(vP)/\partial\phi_i]$. In a stationary distribution $P_\mathrm{stat}$, by definition, the divergence of $\bm{J}$ vanishes, corresponding to incompressible flows. If only one field is present, this means that $J$ is uniform in field space, and that, in most interesting situations, the probability current itself vanishes. For example, if field space is unbounded, the normalisation condition $\int P_\mathrm{stat}\dd\phi = 1$ requires that $P_\mathrm{stat}$ decreases at infinity strictly faster than $\vert \phi \vert^{-1}$. In this case, both $P_\mathrm{stat}(\phi)$ and $\partial P_\mathrm{stat}(\phi)/\partial\phi$ vanish at infinity, hence everywhere, and this gives
\begin{align}
\label{eq:stationaryDistrib}
P_\mathrm{stat}\propto\frac{\ee^{\frac{1}{v}}}{v}\, .
\end{align}
If more than one field is present, one can already see that the analysis is much less trivial. The distribution~(\ref{eq:stationaryDistrib}) is still a solution of the stationarity problem, but non-uniform probability currents are also allowed (since only their divergence must vanish), yielding other solutions. As will be seen, going from one to several fields always renders the problems more difficult, but makes the phenomenology of their solutions richer.
\subsection{First Passage Time Analysis}
\begin{figure}[t]
\begin{center}
\includegraphics[width=0.7\textwidth]{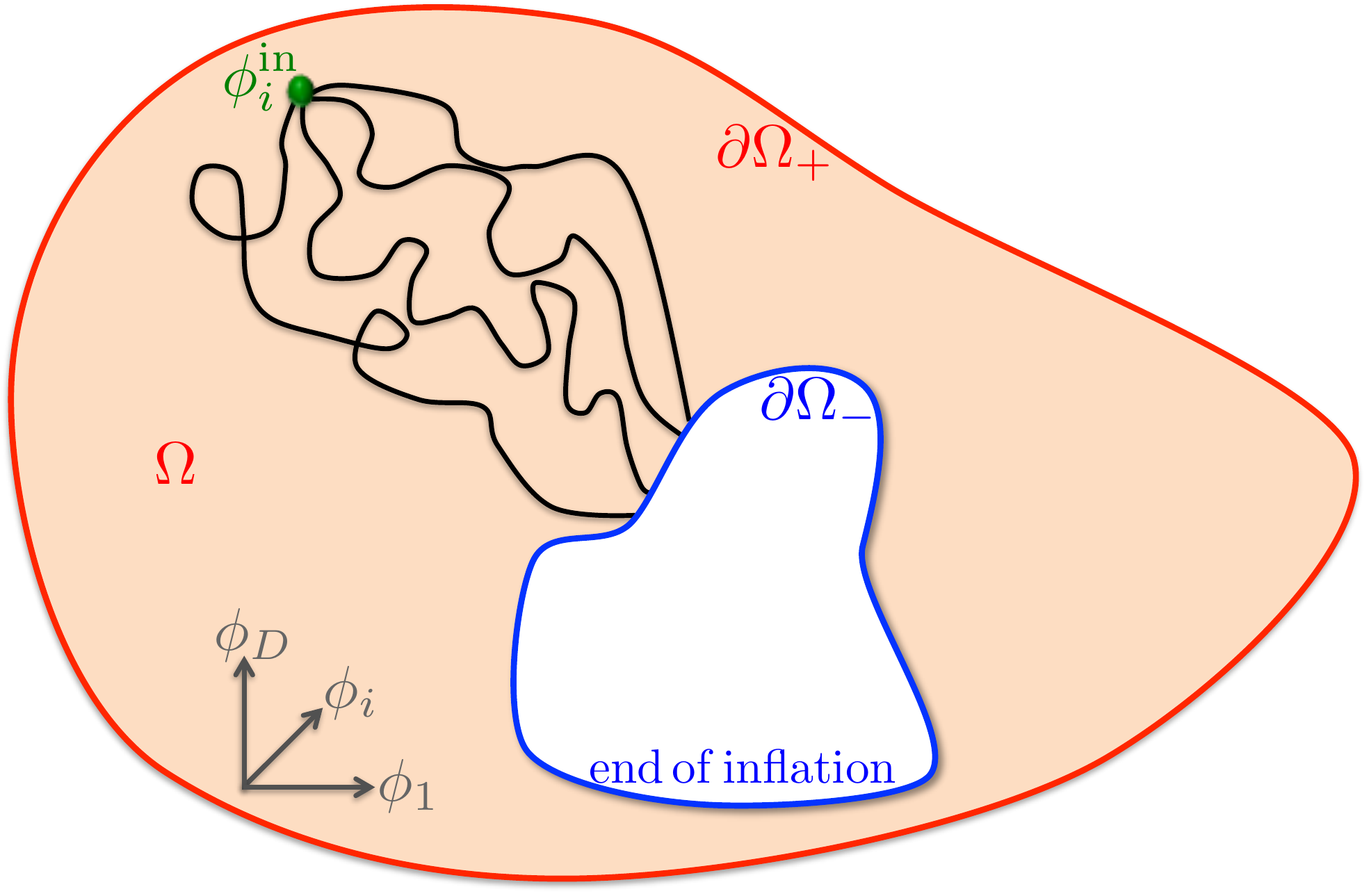}
\caption{First Passage Time problem. Starting from initial field values $\phi_i^\uin$, different trajectories (black lines) realise different numbers of \efolds $\mathcal{N}$ in the inflationary domain $\Omega$ (pale red region) until the final hypersurface $\partial\Omega_-$ (blue line) is reached. Note that $\Omega$ may also be bounded ``from above'' by $\partial\Omega_+$ (red line). The moments of this random variable $\mathcal{N}$ are the solutions of \Eqs{eq:PDE}.}
\label{fig:sketchFirstPassageTime}
\end{center}
\end{figure}
For the reasons explained in \Sec{sec:intro}, we are interested in the duration of the inflationary dynamics described by \Eq{eq:Langevin}, or equivalently, by \Eq{eq:FokkerPlanck}. More precisely, the problem we want to solve is the one depicted in \Fig{fig:sketchFirstPassageTime}. The inflationary domain in field space is denoted $\Omega$. In the standard situation, it is given by the set of points such that the first slow-roll parameter $\epsilon_1=\Mp^2\sum_i v_{\phi_i}^2/v^2$ is less than one, $\Omega=\lbrace (\phi_1,\cdots,\phi_D)\vert \epsilon_1(\phi_1,\cdots,\phi_D)<1\rbrace$. However, in order to also consider situations where inflation does not end by slow-roll violation~\cite{Linde:1993cn,Copeland:1994vg,Bernardeau:2002jf,Bernardeau:2004zz,Lyth:2005qk,Alabidi:2006wa}, in the following, $\Omega$ is left unspecified. Starting from $\phi_i^\mathrm{in}$ at time $N_\uin$,\footnote{We therefore consider the case of an infinitely localised initial distribution $P_\uin(\phi_i)=\delta^D(\phi_i-\phi_i^\uin)$. For more generic initial conditions $P_\uin(\phi_i)$, the result can be obtained by simply averaging the ones derived here with $P_\uin(\phi_i)$.} let $\mathcal{N}$ denote the number of \efolds realised before $\partial\Omega_-$, the boundary of $\Omega$ where inflation stops, is crossed. Obviously, $\mathcal{N}$ is different for each realisation of the stochastic process under consideration. 

In \App{app:FPT} (see also \App{app:FPT:Langevin}), it is shown how the first passage time formalism~\cite{Bachelier:1900, Gihman:1972} allows one to express the moments of $\mathcal{N}$ as the solutions of deterministic differential equations. Introducing the set of functions
\begin{align}
f_n(\phi_i)\equiv \langle \mathcal{N}^n\rangle(\phi_i^\uin=\phi_i)\, ,
\end{align}
it is found that, at leading order in slow roll, they satisfy the hierarchy of partial differential equations\footnote{
For $n\geq 2$, it is sometimes be more convenient to work with the centred moments
\begin{align}
g_n(\phi_i)\equiv\langle \left(\mathcal{N}-\langle\mathcal{N}\rangle\right)^n\rangle(\phi_i^\uin=\phi_i)\, .
\end{align}
They satisfy a similar hierarchy of partial differential equations, namely
\begin{align}
\sum_i\left(
\frac{v_{\phi_i}}{v}\frac{\partial}{\partial\phi_i}-v\frac{\partial^2}{\partial\phi_i^2}\right)g_n=2 n v\sum_i\frac{\partial f_1}{\partial\phi_i}\frac{\partial g_{n-1}}{\partial\phi_i}+ v n (n-1) g_{n-2}\sum_i\left(\frac{\partial f_1}{\partial\phi_i}\right)^2\, ,
\end{align}
valid for $n\geq 2$ if one defines $g_0=1$ and $g_1=0$.
}
\begin{align}
\sum_i\left(v\frac{\partial^2}{\partial\phi_i^2}-
\frac{v_{\phi_i}}{v}\frac{\partial}{\partial\phi_i}\right)f_n= -n\frac{f_{n-1}}{\Mp^2}\, ,
\label{eq:PDE}
\end{align}
see \Eq{eq:PDE:appendix}. This equation is valid for $n\geq 1$, where we have defined $f_0=1$.  It is the main result of this section and its properties will be discussed in details in what follows. At this stage, let us simply notice that if only one field is present, \Eq{eq:PDE} is an ordinary differential equation, the solution of which is given by~\cite{Vennin:2015hra}
\begin{align}
\label{eq:fn:sol:onefield}
f_n(\phi)=n \int_{\bar{\phi}_n}^\phi\frac{\dd x}{\Mp}\int^{\bar{\bar{\phi}}_n}_x\frac{\dd y}{\Mp}\frac{1}{v(y)}\exp\left[\frac{1}{v(y)}-\frac{1}{v(x)}\right]f_{n-1}(y)\, ,
\end{align}
where $\bar{\phi}_n$ and $\bar{\bar{\phi}}_n$ are integration constants (see \Sec{sec:BoundaryConditions}). All moments can therefore be calculated analytically, for any potential $v(\phi)$. When more than one field is present however, \Eq{eq:PDE} is a partial differential equation, for which no generic analytical solution exists. Partial differential equations are notoriously more difficult to study (even at the numerical level) than ordinary differential equations. This is why multiple field scenarios are more challenging from a technical point of view. 
\subsection{Boundary Conditions}
\label{sec:BoundaryConditions}
Single- and multi-field scenarios also differ when considering the boundary conditions according to which \Eq{eq:PDE} must be solved. Such conditions need to be specified, for instance to fix the integration constants $\bar{\phi}_n$ and $\bar{\bar{\phi}}_n$ appearing in \Eq{eq:fn:sol:onefield}. 

A first requirement is that $f_n=0$ on $\partial\Omega_-$, which simply encodes the fact that, if one starts off the evolution on $\partial \Omega_-$, one instantaneously crosses $\partial \Omega_-$ by definition, so that $\mathcal{N}=0$ and all its moments vanish. In case $\Omega$ is compact in field space (as in hilltop models) so that the inflationary domain is bounded by $\partial\Omega_-$ in all directions, this defines a Cauchy problem (\ie there exists a single solution once the boundary condition $\left. f_n\right \vert_{\partial\Omega_-}=0$ is imposed). In all other situations however, where inflation can proceed at arbitrarily large field value (see \Fig{fig:sketchFirstPassageTime}) as in large-field or plateau models, this single absorbing condition on $\partial\Omega_-$ is not sufficient and one needs to add a second boundary condition. This is why in what follows, a second reflecting (or absorbing) boundary condition is placed at large field value on $\partial\Omega_+$. 

The impact of this extra boundary condition is in fact very much dependent on whether one or several fields are present. Indeed, in \App{app:FPB}, it is shown that the probability $p_+(\phi_i)$ to bounce again (or being absorbed into) the extra boundary $\partial\Omega_+$ before exiting inflation through $\partial\Omega_-$, starting from $\phi_i^\uin=\phi_i$, is given by the solution of the ordinary differential equation
\begin{align}
\sum_i\left(v\frac{\partial^2}{\partial\phi_i^2}-
\frac{v_{\phi_i}}{v}\frac{\partial}{\partial\phi_i}\right)p_+=0\, ,
\label{eq:FPB:equadiff}
\end{align}
with boundary conditions $p_+=1$ on $\partial\Omega_+$ and $p_+=0$ on $\partial\Omega_-$, see \Eq{eq:FPB:equadiff:app}. If only one scalar field is present, the solution of this equation is given by~\cite{Vennin:2015hra}
\begin{align}
p_+\left(\phi_\uin\right)=\dfrac{\displaystyle\int_{\phi_-}^{\phi_\uin}\ee^{-\frac{1}{v\left(\phi\right)}}\dd \phi}{\displaystyle\int_{\phi_-}^{\phi_+}\ee^{-\frac{1}{v\left(\phi\right)}}\dd \phi}\, ,
\end{align}
where one has noted $\partial\Omega_\pm=\lbrace \phi_\pm\rbrace$. From this expression, one can see that as soon as $v$ is not vanishing at large-field value [or if it does, as soon as it does not go to $0$ faster than $1/\ln(\phi)$], $p_+\rightarrow 0$ when $\phi_+\rightarrow\infty$. This means that the system never (\ie with probability $0$) explores large-field regions of the potential. As will be shown in \Sec{sec:InfiniteInflation}, this implies that, in most single-field setups, the exact location of the second boundary cancels out in all physical results when removed to infinity. In \Sec{sec:wallproba} however, it will be shown that this property does not generalise to multi-field scenarios. This is one of the reasons why different results can be obtained in these setups.
\section{Harmonic Potentials}
\label{sec:harmonic}
In the previous section, the partial differential equations~(\ref{eq:PDE}) governing the moments of the number of inflationary \efolds were derived. For a fully generic multi-field potential, these equations have no analytical solutions and one needs to resort to numerical analysis. Alternatively, in this section we identify a subclass of inflationary potentials for which \Eq{eq:PDE} can be solved exactly and the effects associated with the inclusion of multiple fields can be studied analytically.
\subsection{Polar Coordinates}
Let us first note that \Eqs{eq:PDE} are diffusion equations, akin to the Laplace equation. This suggests that some insight may be gained by reparameterising field space with polar-type coordinates. Since the slow-roll trajectory follows the gradient of the potential at the classical level [\ie without including the diffusion term in \Eq{eq:Langevin}], a natural choice\footnote{This choice of coordinates is also similar to the adiabatic-entropic decomposition of \Ref{Gordon:2000hv}, if one interprets Eqs.~(31),~(32) and~(35) of this reference by replacing the field derivatives by their classical equations of motion.} is to take the potential $v$ itself for the radial coordinate, completed by $D-1$ angular coordinates $\theta_j$ (with $1 \leq j \leq D-1$).\footnote{Strictly speaking, this procedure is well-defined only if the level lines of $v(\phi_i)$ form simply connected hyper surfaces in field space. This is implicitly assumed in what follows, even if more complicated situations can also be studied, either making use of symmetries in the potential function $v(\phi_i)$ as in hybrid inflation, or paving field space with several maps.} By expanding $\partial/\partial\phi_i = v_{\phi_i} \partial/\partial v + \sum_j (\theta_j)_{\phi_i}\partial/\partial\theta_j$ in the new coordinates system, the differential operator of \Eq{eq:PDE} can be written as 
\begin{align}
&\sum_{i=1}^D\left(
v\frac{\partial^2}{\partial\phi_i^2}-\frac{v_{\phi_i}}{v}\frac{\partial}{\partial\phi_i}\right)= 
v \left\vert  \bm{\nabla}(v)\right\vert^2 \frac{\partial^2}{\partial v^2}
+v\sum_{j,\ell = 1}^{D-1} 
\bm{\nabla}(\theta_j)\cdot \bm{\nabla}(\theta_\ell)
\frac{\partial^2}{\partial\theta_j\partial\theta_\ell}
\nonumber\\ 
&+2 v\sum_{j=1}^{D-1}
\bm{\nabla}(\theta_j)\cdot \bm{\nabla}(v)
\frac{\partial^2}{\partial v \partial\theta_j}
+\left[v \Delta v-\frac{1}{v} \left\vert  \bm{\nabla}(v)\right\vert^2\right]\frac{\partial}{\partial v}
+\sum_{j=1}^{D-1}\left[v\Delta\theta_j-\frac{1}{v}\bm{\nabla}(\theta_j)\cdot \bm{\nabla}(v)\right]\frac{\partial}{\partial\theta_j}\, .
\label{eq:Fpop:radial:1}
\end{align}
In this expression, recall that the vectorial notation (and the differential operators $\bm{\nabla}$ and $\Delta=\vert\bm{\nabla}^2\vert$) refer to field space. For example, $\bm{\nabla}(v)=\sum_{i=1}^D v_{\phi_i} \bm{\mathrm{e}}_{\phi_i}$, where $\lbrace \bm{\mathrm{e}}_{\phi_i}\rbrace $ stands for the field space basis. One can always choose the angular variables $\theta_j$ to form a system of orthogonal variables\footnote{For example~\cite{Malik:1998gy}, one can start from $\bm{\nabla}(v)$ and use Gram-Schmidt orthogonalisation procedure to iteratively derive $\bm{\nabla}(\theta_1)$, $\bm{\nabla}(\theta_2)$, etc.} and one obtains
\begin{align}
& \sum_{i=1}^D\left(
v\frac{\partial^2}{\partial\phi_i^2}-\frac{v_{\phi_i}}{v}\frac{\partial}{\partial\phi_i}\right)= 
\nonumber\\ & \quad\quad
v \left\vert  \bm{\nabla}(v)\right\vert^2 
\left\lbrace
\frac{\partial^2}{\partial v^2}
+\sum_{j = 1}^{D-1} 
\frac{\left\vert  \bm{\nabla}(\theta_j)\right\vert^2}{\left\vert  \bm{\nabla}(v)\right\vert^2}
\frac{\partial^2}{\partial\theta_j^2}
+\left[\frac{\Delta v}{\left\vert  \bm{\nabla}(v)\right\vert^2}-\frac{1}{v^2} \right]\frac{\partial}{\partial v}
+\sum_{j=1}^{D-1}\frac{\Delta\theta_j}{\left\vert  \bm{\nabla}(v)\right\vert^2} \frac{\partial}{\partial\theta_j}
\right\rbrace \, .
\label{eq:Fpop:radial}
\end{align}
\subsection{Harmonic Potentials}
\label{subsec:harmonic}
We now restrict the analysis to potentials for which separable solutions (in the basis $\lbrace v,\,\theta_j\rbrace$) of \Eqs{eq:PDE} exist. An important remark is that purely radial (\ie independent of $\theta_j$) solutions of \Eqs{eq:PDE} can be found if the coefficient in front of $\partial/\partial v$ is a function of $v$ only. For this reason, we define ``harmonic potentials'' as being such that 
\begin{align}
g\equiv  \frac{\Delta v}{ \left\vert  \bm{\nabla}(v)\right\vert^2}
\label{eq:harmonic:g}
\end{align}
is a function of $v$ only.

In order to understand to which extent harmonic potentials allow one to proceed analytically, let us discuss the case where $D=2$ fields are present. For two-field potentials, one has a single angular variable $\theta$, and the orthogonality condition $\bm{\nabla}(v)\perp \bm{\nabla}(\theta)$ mentioned between \Eqs{eq:Fpop:radial:1} and~(\ref{eq:Fpop:radial}) implies that $\bm{\nabla}(\theta)=h (- v_{\phi_2}\bm{\mathrm{e}}_{\phi_1}+  v_{\phi_1}\bm{\mathrm{e}}_{\phi_2})$, where $h$ is an overall factor that is left unspecified at this stage. Let us simply note that, in order for $\theta$ to be globally defined~\cite{Saffin:2012et}, the curl of $\bm{\nabla}(\theta)$ must vanish, $\theta_{\phi_1\phi_2}=\theta_{\phi_2\phi_1}$, which translates into $ h_{\phi_1} v_{\phi_1} + h_{\phi_2} v_{\phi_2} +h(v_{\phi_1\phi_1}+v_{\phi_2\phi_2})=0$. This is the only condition $h$ needs to satisfy, and for harmonic potentials where $g$ depends on $v$ only, it is interesting to notice that it can be fulfilled if $h$ is taken as depending on $v$ only as well, according to\footnote{Indeed, in this case, one has $h_{\phi_i}=v_{\phi_i}\dd h/\dd v = -g v_{\phi_i} h $ and one can easily check that $ h_{\phi_1} v_{\phi_1} + h_{\phi_2} v_{\phi_2} +h(v_{\phi_1\phi_1}+v_{\phi_2\phi_2})=0$.} $h(v)=\exp[-\int^v g(v')\dd v']$. One can also check that in this case, $\Delta\theta = h_{\phi_2} v_{\phi_1}-h_{\phi_1} v_{\phi_2}=0$. The differential operator of \Eq{eq:PDE} then takes the simple form
\begin{align}
\sum_{i=1}^2\left(
v\frac{\partial^2}{\partial\phi_i^2}-\frac{v_{\phi_i}}{v}\frac{\partial}{\partial\phi_i}\right)= &
v \left\vert  \bm{\nabla}(v)\right\vert^2 
\left\lbrace
\frac{\partial^2}{\partial v^2}
+
\exp\left[-2\!\!\int^v\!\!\!g(v')\dd v'\right]
\frac{\partial^2}{\partial\theta^2}
+\left[g(v)-\frac{1}{v^2} \right]\frac{\partial}{\partial v}
\right\rbrace \, .
\label{eq:Fpop:radial:2D}
\end{align}
In this case, up to the overall $\vert  \bm{\nabla}(v)\vert^2$ factor, all coefficients of the differential operator of \Eq{eq:PDE}  are explicit functions of the radial coordinate $v$ only, and the problem boils down to solving ordinary differential equations after Fourier transforming the angular coordinate, as will be exemplified in \Secs{sec:InfiniteInflation} and~\ref{sec:InhomogeneousEnd}. Let us give a few concrete examples of harmonic potentials.
\subsection{$v(r)$ Potentials}
\label{sec:v(r)}
A subclass of harmonic potentials is provided by potentials $v(r)$ that depend on
\begin{align}
r^2\equiv\sum_i\phi_i^2
\end{align} 
only. Indeed, in this case, one can show that $g=(D-1)/[rv^\prime(r)]+v^{\prime\prime}(r)/{v^\prime}^2(r)$ depends on $r$, hence on $v$, only. In fact, $ \left\vert  \bm{\nabla}(v)\right\vert^2 = {v^\prime}^2(r)$ depends on $v$ only as well. The angular coordinates can be chosen to match the ones of the usual spherical coordinates system in $D$ dimension, given in \App{app:spericalCoord} where their gradients and Laplacians are also derived. This gives rise to
\begin{align}
\sum_{i=1}^D\left(
v\frac{\partial^2}{\partial\phi_i^2}-\frac{v_{\phi_i}}{v}\frac{\partial}{\partial\phi_i}\right)= &
v(r)
\left\lbrace
\frac{\partial^2}{\partial r^2}
+\sum_{j = 1}^{D-1} 
\left[r\displaystyle\prod_{\ell=1}^{j-1}\sin(\theta_\ell)\right]^{-2}
\frac{\partial^2}{\partial\theta_j^2}
\right. \nonumber\\ & \left. 
+\left[\frac{D-1}{r} - \frac{v^\prime(r)}{v^2(r)} \right]\frac{\partial}{\partial r}
+\sum_{j=1}^{D-1}\frac{D-1-j}{\tan(\theta_j)}\left[r\displaystyle\prod_{\ell=1}^{j-1}\sin(\theta_\ell)\right]^{-2}\frac{\partial}{\partial\theta_j}
\right\rbrace \, ,
\label{eq:Fpop:radial:v(r)}
\end{align}
where, for simplicity, $r$ has been used as the radial coordinate instead of $v$. It is interesting to notice that, compared to the single-field case where $D=1$, angular terms involving $\partial/\partial\theta_j$ and $\partial^2/\partial\theta_j^2$ are obviously present, but the radial term proportional to $\partial/\partial r$ also receives a new contribution. One can also check that when $D=2$, \Eq{eq:Fpop:radial:2D} is recovered. These $v(r)$ potentials are further studied in \Sec{sec:InfiniteInflation}.
\subsection{Linear Potentials}
\label{sec:linear}
Another subclass of harmonic potentials is provided by potentials $v(u)$ that depend on a linear combination of the fields
\begin{align}
u=\sum_i\alpha_i\phi_i
\end{align}
only. Here, one can choose the $\alpha_i$ constants to be normalised so that $\sum\alpha_i^2=1$. In this case, one has $\vert\bm{\nabla}(v)\vert^2={v^\prime}^2(u)$ and $\Delta v=v^{\prime\prime}(u)$, so that $g=v^{\prime\prime}(u)/{v^\prime}^2(u)$. The ``angular'' coordinates can then be defined with constant gradients so that $\lbrace \bm{\alpha},\bm{\nabla}(\theta_j)\rbrace$ form an orthonormal basis of field space (here, the $\theta_j$ variables are unbounded and should not be viewed as geometrical angles, and ``angular'' must be understood in a generic way). In this case, one obtains
\begin{align}
\sum_{i=1}^D\left(
v\frac{\partial^2}{\partial\phi_i^2}-\frac{v_{\phi_i}}{v}\frac{\partial}{\partial\phi_i}\right)= &
v \frac{\partial^2}{\partial u^2}
+v \sum_{j = 1}^{D-1} 
\frac{\partial^2}{\partial\theta_j^2}
-\frac{v^\prime(u)}{v(u)}\frac{\partial}{\partial u}
 \, ,
\label{eq:Fpop:linear}
\end{align}
where, for simplicity, $u$ has been used as the radial coordinate instead of $v$. From this expression, it is clear that the situation is very close to a single-field setup, since inflation is only driven by the ``scalar field'' $u$. The only difference with a purely single-field setup arises if the boundary conditions discussed in \Sec{sec:BoundaryConditions} depend on the other fields, and introduce some ``angular'' dependence in the solutions of \Eqs{eq:PDE}. This situation is further investigated in \Sec{sec:InhomogeneousEnd}.
\subsection{Straight Potentials}
\label{sec:straight}
\begin{figure}[t]
\begin{center}
\includegraphics[width=0.5\textwidth]{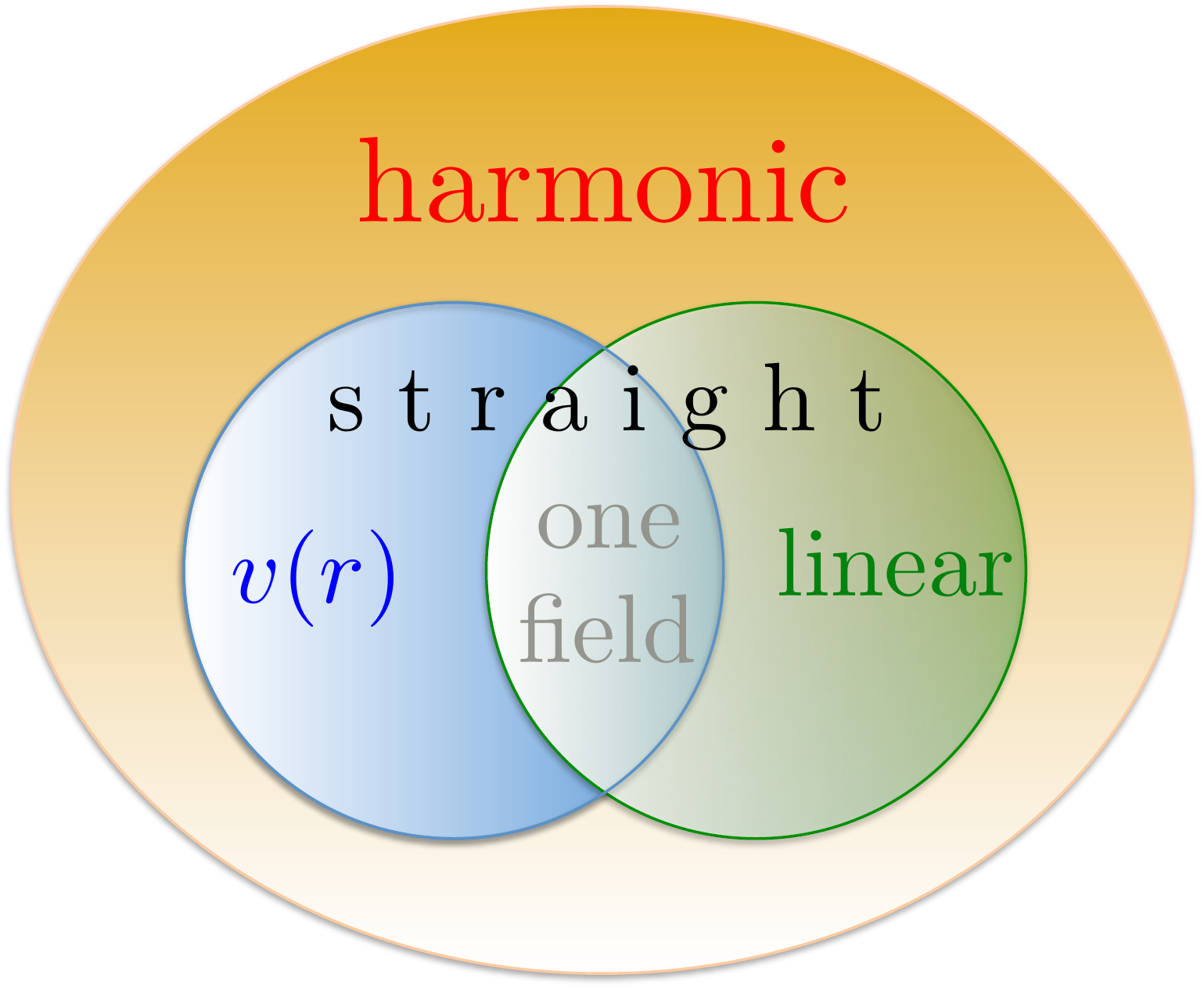}
\caption{``Harmonic Potentials'' (see \Sec{subsec:harmonic}) are defined through the condition~(\ref{eq:harmonic:g}) that $\Delta v/\vert \bm{\nabla}(v)\vert^2$ is a function of $v$ only and are such that the statistical moments of the number of inflationary \efolds can be worked out analytically. A specific class of harmonic potentials is provided by ``straight potentials'' (see \Sec{sec:straight}) for which the slow-roll classical trajectories are straight lines in field space. Those are made of ``linear potentials'' (see \Sec{sec:linear}), \ie potentials that depend on a linear combination of the fields only, and ``$v(r)$ potentials'' (see \Sec{sec:v(r)}), \ie potentials that depend on $r^2=\sum\phi_i^2$ only. Single-field potentials lie at the intersection between these two.}
\label{fig:sketchHarmonic}
\end{center}
\end{figure}
In \Secs{sec:v(r)} and~\ref{sec:linear}, we showed that $v(r)$ and linear potentials are ``harmonic'' in the sense defined in \Sec{subsec:harmonic}. In fact, as we are now going to see, such potentials share the property that the slow-roll classical trajectories are straight lines in field space. We call such potentials ``straight potentials''. On large scales, entropy perturbations can source adiabatic perturbations only if the background solution follows a curved trajectory in field space~\cite{Gordon:2000hv}, which is why these potentials are exactly the ones for which adiabatic perturbations are conserved on large scales, at least at the classical level.  We therefore expect them to play a special role in the present analysis. For this reason, in this section, we try to better characterise them.

Since the slow-roll classical trajectories follow the local gradients of the potential, starting from some point $\bm{\phi}$, the next point on the classical trajectory has coordinates $\bm{\phi}+\epsilon\,\bm{\nabla}(v)$, $\epsilon$ being an infinitesimal number. The gradients evaluated at these two points must be parallel for straight potentials. In other words, the variation in the gradients between $\bm{\phi}$ and $\bm{\phi}+\epsilon\,\bm{\nabla}(v)$ must be aligned with the gradient at $\bm{\phi}$, that is to say
\begin{align}
\left[{\bm{H}}(v)\cdot \bm{\nabla}(v)\right] \wedge \bm{\nabla}(v) =\bm{0}\, ,
\end{align}
where ${\bm{H}}(v)=\bm{\nabla}^2(v)=\sum_{i,k}\partial^2 v/(\partial\phi_i\partial\phi_k)\bm{e}_{\phi_i} \otimes \bm{e}_{\phi_k}$ is the Hessian matrix of $v$. By expanding this relation into its components $\bm{e}_{\phi_i}$, it is easy to show that it leads to
\begin{align}
\left\lbrace \left\vert \bm{\nabla}(v) \right\vert^2,v\right\rbrace_{\phi_i,\phi_{i+1}}=0
\label{eq:grad(v):PB}
\end{align}
for all $1\leq i < D$, where $\left\lbrace  a,b \right\rbrace_{\phi_i,\phi_{i+1}}\equiv a_{\phi_i} b_{\phi_{i+1}} - a_{\phi_{i+1}} b_{\phi_{i}}$ stands for the $i^\mathrm{th}$ Poisson bracket between $a$ and $b$. Straight potentials are therefore such that all Poisson brackets between $ \left\vert \bm{\nabla}(v) \right\vert$ and $v$ vanish, meaning that  $ \left\vert \bm{\nabla}(v) \right\vert$ depends on $v$ only.\footnote{
A first remark is that if the Poisson brackets $\lbrace , \rbrace_{\phi_{i},\phi_{i+1}}$ vanish, all Poisson brackets vanish. For example, it is easy to show that
\begin{align}
\frac{\partial b}{\partial\phi_2}\left\lbrace a,b\right\rbrace_{\phi_1,\phi_3} = \frac{\partial b}{\partial\phi_3}\left\lbrace a,b\right\rbrace_{\phi_1,\phi_2} + \frac{\partial b}{\partial\phi_1}\left\lbrace a,b\right\rbrace_{\phi_2,\phi_3}
\end{align}
so that if $\left\lbrace a,b\right\rbrace_{\phi_1,\phi_2} = \left\lbrace a,b\right\rbrace_{\phi_2,\phi_3}=0$, then $\left\lbrace a,b\right\rbrace_{\phi_1,\phi_3}=0$, so on and so forth. Then, in the basis $\lbrace v,\theta_1,\cdots \theta_{D-1}\rbrace$, the Poisson bracket $\lbrace a,b \rbrace_{v,\theta_j}$ is given by
\begin{align}
\lbrace a,b \rbrace_{v,\theta_j} &= \frac{\partial a}{\partial v}\frac{\partial b}{\partial \theta_j} -  \frac{\partial a}{\partial \theta_j}\frac{\partial b}{\partial v} 
= \sum_{i,k}  \frac{\partial a}{\partial \phi_i}\frac{\partial\phi_i}{\partial v}\frac{\partial b}{\partial\phi_k}\frac{\partial\phi_k}{\partial \theta_j}
 - \sum_{i,k}   \frac{\partial a}{\partial\phi_i}\frac{\partial\phi_i}{\partial \theta_j}\frac{\partial b}{\partial\phi_k}\frac{\partial\phi_k}{\partial v}\\
 &=\sum_{i,k} \frac{\partial\phi_i}{\partial v}\frac{\partial\phi_k}{\partial \theta_j} \left( \frac{\partial a}{\partial \phi_i}\frac{\partial b}{\partial\phi_k} - \frac{\partial a}{\partial \phi_k}\frac{\partial b}{\partial\phi_i}\right)
 =\sum_{i,k} \frac{\partial\phi_i}{\partial v}\frac{\partial\phi_k}{\partial \theta_j} \left\lbrace a,b\right\rbrace_{\phi_i,\phi_k}\, .
\end{align} 
Therefore, if all Poisson brackets between $a$ and $b$ vanish, then $\lbrace a,b \rbrace_{v,\theta_j} $ vanishes as well. If one takes $b=v$, this means that $\partial a/\partial\theta_j=0$, for all $1\leq j \leq D-1$, hence $a$ depends on $v$ only, as is the case for $ \left\vert \bm{\nabla}(v) \right\vert$ in \Eq{eq:grad(v):PB}.
\label{footnote:Poisson}
}
Since this quantity appears in various places in \Eq{eq:Fpop:radial}, we understand why straight potentials play a special role in the present context. In particular, in \Sec{sec:v(r)}, it was shown that for $v(r)$ potentials, $\left\vert \bm{\nabla}(v) \right\vert^2 = {v^\prime}^2(r)$ is a function of $v$ only, which confirms that $v(r)$ potentials are straight potentials. Similarly in \Sec{sec:linear}, it was shown that for linear potentials, $\left\vert \bm{\nabla}(v) \right\vert^2 = {v^\prime}^2(u)$ is a function of $v$ only, and linear potentials also are straight potentials, as announced above.


Reciprocally, one can show that straight potentials can only be of one of these two types: $v(r)$ potentials or linear potentials. Indeed, let us consider a straight potential $v$ and its (straight) gradient lines in dimension $D=2$. We first assume that its gradient lines never intersect in field space. This means that they all are parallel, and one can write $\bm{\nabla}(v) = a(\bm{\phi})\sum_i\alpha_i\bm{e}_{\phi_i}$, hence $v_{\phi_i}=a(\bm{\phi})\alpha_i$. One then has $\lbrace v, \sum_i\alpha_i\phi_i \rbrace_{\phi_k,\phi_\ell} = \alpha_\ell v_{\phi_k} - \alpha_k v_{\phi_\ell} = 0$, hence $v$ depends on $\sum_i\alpha_i\phi_i$ only (see footnote~\ref{footnote:Poisson}) and is therefore linear. Let us now assume that there is exactly one intersection point in the gradient lines of $v$, which, after performing a constant field shift, we set at the origin of field space. It is easy to see that any gradient line not passing through the origin would produce a second intersection point at least, hence all gradient lines go through the origin and one can write $\bm{\nabla}(v)=a(\bm{\phi}) \bm{e}_r$, where $\bm{e}_r$ is the unit vector pointing to the radial direction $r=\sqrt{\sum\phi_i^2}$. This means that $v_{\phi_i}=a(\bm{\phi})\phi_i/r$. Since $r_{\phi_i}=\phi_i/r$, one has $\lbrace v,r\rbrace_{\phi_k,\phi_\ell}= v_{\phi_k} r_{\phi_\ell} - v_{\phi_\ell} r_{\phi_k}=0$, hence $v$ depends on $r$ only and is of the $v(r)$ type. Finally, let us assume that the gradient lines of $v$ intersect at two or more points. Then, one can convince oneself that an infinite number of other intersection points can be obtained, that fill the entire (or a dense subset of the) field space. Since the gradient of $v$ must vanish when two non-parallel lines intersect (otherwise its direction would be ill-defined), this means that $v$ is constant, and this case is in fact trivial. This result can be generalised to $D>2$ where one finds that the potential is of the $v(r)$ type within the field subspace that is orthogonal to the one containing the fields of which $v$ is independent.

The situation is schematically summarised in \Fig{fig:sketchHarmonic}. Straight potentials are a specific class of harmonic potentials. They are either linear or $v(r)$ potentials, and single-field potentials lie at the intersection between these two. Let us finally notice that not all harmonic potentials are straight. For example, let us consider a ``loop corrected'' potential of the form $v=v_0[1+\alpha\sum_{i=1}^D\log\left(\phi_i/\Mp\right)]$. The function $g$ defined in \Eq{eq:harmonic:g} is constant, $g=-1/(v_0\alpha)$, and such potentials are therefore harmonic. However, one has $\lbrace\vert \bm{\nabla}(v)\vert^2 , v \rbrace_{\phi_k,\phi_\ell} = 2 / (v_0 \alpha \phi_k \phi_\ell) (1 / \phi_\ell^2 - 1 / \phi_k^2)$ which is not a vanishing function, hence loop corrected potentials are not straight. More generally, this is the case for all potentials $v(w)$ that are functions of $w=\prod_i\phi_i$ only, for which $g=v''(w)/{v'}^2(w)$.
\section{$v(r)$ Potentials and Infinite Inflation}
\label{sec:InfiniteInflation}
\begin{figure}[t]
\begin{center}
\includegraphics[width=0.49\textwidth]{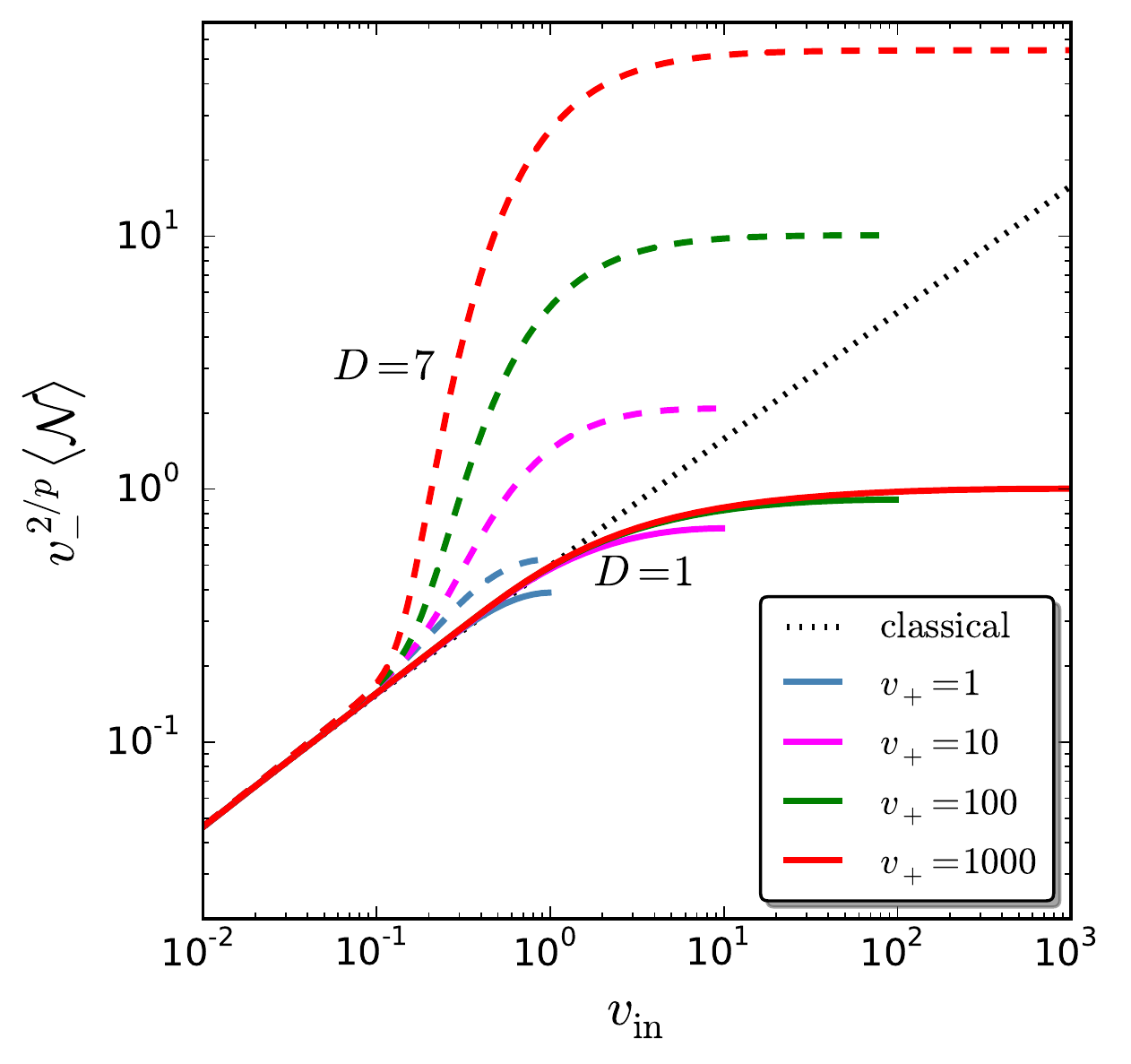}
\includegraphics[width=0.478\textwidth]{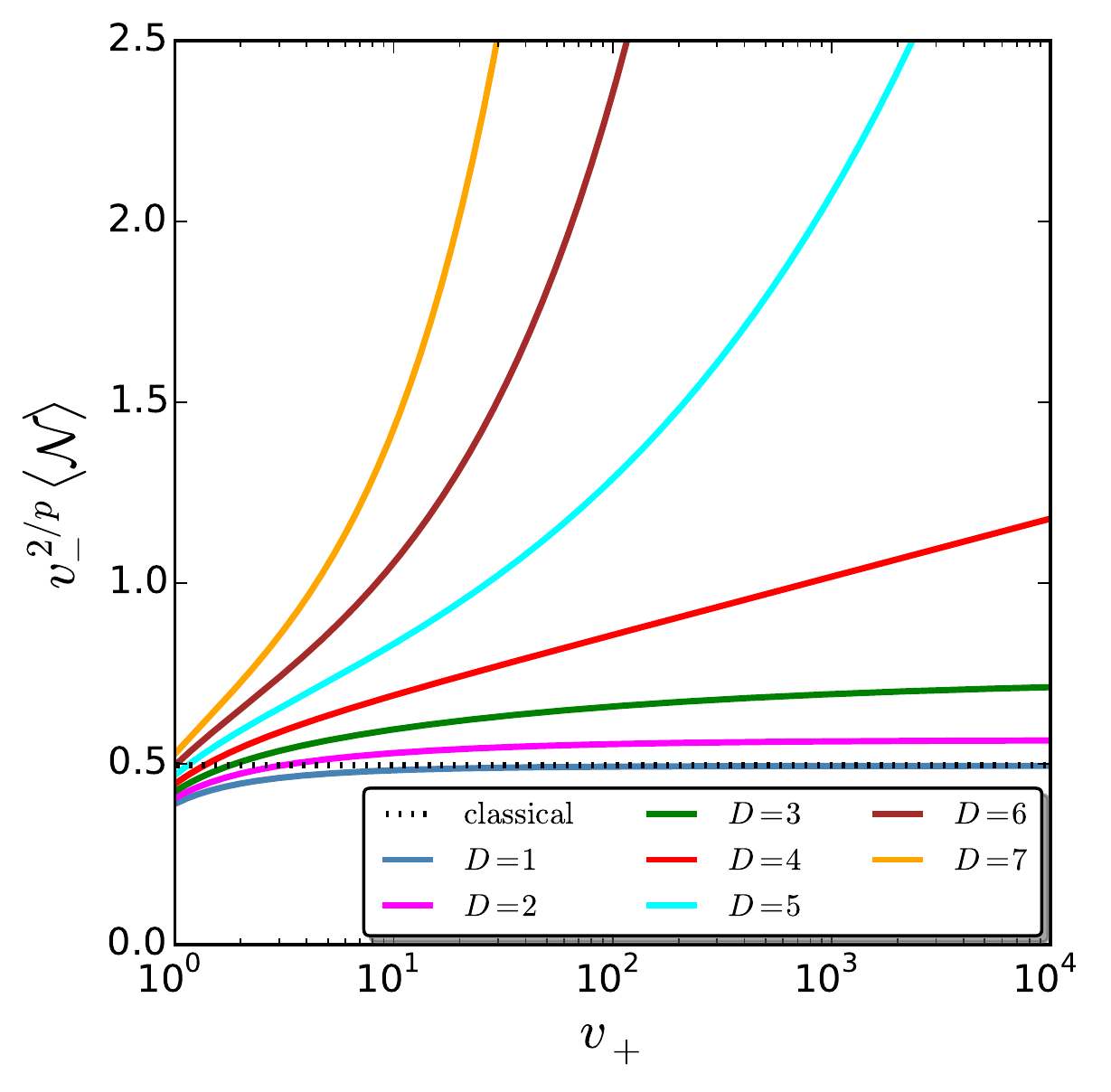}
\caption{Mean number of \efolds(\ref{eq:v(r):sol}) for monomial potentials $v(r)\propto r^p$ with $p=4$ (rescaled by $v_-^{2/p}$, with $r_-=p/\sqrt{2}$ where inflation ends by slow-roll violation). In the left panel, $\langle\mathcal{N}\rangle$ is displayed as a function of the initial condition $v_\uin$, for $D=1$ (solid lines) and $D=7$ (dashed lines) and for different values of $v_+>v_\uin$ at which a reflecting boundary condition is placed. The black dotted line stands for the classical limit~(\ref{v(r):Ncl}), $N_\ucl$, towards which the stochastic results asymptote when $v\ll 1$. In the opposite regime, $\langle \mathcal{N} \rangle$ deviates from $N_\ucl$ in a way that depends on $v_+$ and $D$, and that is further discussed in the main text. One should note that, in principle, $v_\uin>1$ lies outside the validity range of the present calculation since it corresponds to initial super-Planckian energy density, but it is displayed to make clear the asymptotic behaviour of the mean number of \efolds at large initial field value. In the right panel, $v_\uin=1$ is fixed, but $v_+$ varies, and different values of $D$ are displayed. One can check that when $D < p$, a finite asymptotic value is reached when $v_+\rightarrow\infty$, while when $D \geq p$, $\langle N\rangle$ diverges in this limit.}
\label{fig:Nmean:v(r)}
\end{center}
\end{figure}
In \Sec{sec:v(r)}, it was shown that $v(r)$ potentials provide a subclass of harmonic potentials, for which analytical solutions of the diffusion equations~(\ref{eq:PDE}) can be found. In this section, we derive such solutions and use these potentials to illustrate the physical implications of including more than one scalar field in the analysis.

If one sets boundary conditions to be angular independent, that is to say if one assumes that inflation ends at $r=r_-$ and that a reflecting wall is placed at $r=r_+$, angular independent solutions of \Eq{eq:PDE} can be obtained. More precisely, combining \Eqs{eq:PDE} and~(\ref{eq:Fpop:radial:v(r)}), one obtains
\begin{align}
f_n(r)=
n\displaystyle\int_{r_-}^r\frac{\dd r^\prime}{\Mp}
\displaystyle\int_{r^\prime}^{r_+}\frac{\dd r^{\prime\prime}}{\Mp}
\dfrac{\ee^{\frac{1}{v(r^{\prime\prime})}-\frac{1}{v(r^{\prime})}}}{v(r^{\prime\prime})}\left(\frac{r^{\prime\prime}}{r^\prime}\right)^{D-1}f_{n-1}(r^{\prime\prime})
\label{eq:v(r):sol}
\end{align}
where we recall that $f_0=1$. One can check that $f_n(r_-)=0$ (absorbing boundary condition) and that $f^\prime(r_+)=0$ (reflecting boundary condition). If one wanted to place an absorbing boundary condition at $r_+$ instead, one would have to change the upper bound of the second integral to a smaller value~\cite{Vennin:2015hra}, but this would not affect the following considerations. In the same manner, if, instead of the situation depicted in \Fig{fig:sketchFirstPassageTime} where the fields classically decrease during inflation, one considered a hilltop potential symmetric about $r=0$, one would have to replace $r_+$ by $0$ in the above formula, and this case is also discussed in what follows. 

A preliminary remark is that both the number of fields $D$ and the location of the reflecting boundary condition $r_+$ explicitly appear in \Eq{eq:v(r):sol}, and are therefore expected to play a role. For illustration, in the left panel of \Fig{fig:Nmean:v(r)}, the mean number of \efolds(\ref{eq:v(r):sol}), $\langle\mathcal{N}\rangle=f_1(r_\uin)$, is displayed as a function of the initial condition $v_\uin$ for a quartic potential $v\propto r^4$, for different values of $r_+$, and for $D=1$ (solid lines) and $D=7$ (dashed lines). One can check that, in some regimes at least, the result strongly depends on $r_+$ and $D$ indeed (see also the right panel of \Fig{fig:Nmean:v(r)}), in a way that we now analyse in more details.
\subsection{Classical Limit}
\label{sec:v(r):ClassicalLimit}
A first important consistency check consists of verifying that the correct classical limit is recovered. In the classical picture (\ie without the stochastic terms), the slow-roll equation of motion~(\ref{eq:Langevin}) for the fields is given by $\dd \phi_i/\dd N = -\Mp^2v^\prime\phi_i/(v r)$. This gives rise to $\dd r/\dd N=-\Mp^2 v^\prime/v$, and the classical number of \efolds reads
\begin{align}
N_\ucl = \frac{1}{\Mp^2}\int_{r_-}^{r}\frac{v(r^\prime)}{v^\prime(r^\prime)}\dd r^\prime\, .
\label{v(r):Ncl}
\end{align}
Let us see how this formula can be recovered from \Eq{eq:v(r):sol}. In the classical limit, energy densities are sub-Planckian $v\ll 1$ and the integral over $r^{\prime\prime}$ in \Eq{eq:v(r):sol} is dominated by its contribution close to the lower bound $r_-$, near which one can Taylor expand $1/v$ at first order, $1/v(r^{\prime\prime})\simeq  1/v(r^{\prime}) - v^\prime(r^\prime)/v^2(r^\prime) (r^{\prime\prime}-r^\prime)$. This gives rise to
\begin{align}
\displaystyle\int_{r^\prime}^{r_+}{\dd r^{\prime\prime}}
\dfrac{\ee^{\frac{1}{v(r^{\prime\prime})}}}{v(r^{\prime\prime})}\left(r^{\prime\prime}\right)^{D-1}
\simeq 
\ee^{\frac{1}{v(r^\prime)}}\displaystyle\int_{r^\prime}^{r_+}\frac{\dd r^{\prime\prime}}{\Mp}
\left[\frac{1}{v(r^\prime)}-\frac{v^\prime(r^\prime)}{v^2(r^\prime)}\left(r^{\prime\prime}-r^\prime\right)\right]
\ee^{-\frac{v^\prime(r^\prime)}{v^2(r^\prime)}(r^{\prime\prime}-r^\prime)}
\left(r^{\prime\prime}\right)^{D-1}\, .
\end{align}
This integral can be performed through $D-2$ integrations by parts. If one keeps contributions from the upper bound of the integral only and expands the result at leading order in $v$, one obtains
\begin{align}
\displaystyle\int_{r^\prime}^{r_+}{\dd r^{\prime\prime}}
\dfrac{\ee^{\frac{1}{v(r^{\prime\prime})}}}{v(r^{\prime\prime})}\left(r^{\prime\prime}\right)^{D-1}
\simeq \frac{v(r^\prime)}{v^\prime(r^\prime)}\ee^{\frac{1}{v(r^\prime)}}\left(r^\prime\right)^{D-1}\, .
\end{align}
By plugging this formula into \Eq{eq:v(r):sol}, one obtains
\begin{align}
f_1(r)=\langle \mathcal{N}\rangle \simeq N_\ucl
\end{align}
in the classical limit. In the left panel of \Fig{fig:Nmean:v(r)}, the classical formula~(\ref{v(r):Ncl}) is displayed and one can check that, when $v_\uin\ll 1$, it provides a good approximation to the full stochastic results indeed.

The fact that the classical trajectory arises as a saddle-point limit of the full quantum dynamics is not so surprising, since it is  a common feature of path integral calculations in quantum field theory. Let us also notice that the classical limit~(\ref{v(r):Ncl}) depends neither on the number of fields $D$ nor on the location of the upper boundary condition $r_+$, and matches the single-field result. However, as we are now going to see, stochastic corrections break this classical $D$ and $r_+$ invariance and introduce dependence on both the number of fields and the location of the upper boundary condition.
\subsection{Infinite Inflation}
\label{sec:v(r):InfiniteInflation}
\begin{center}
\begin{table}[t]
\definecolor{grey}{RGB}{50,50,50}
\centering
\begin{tabular}{|c|c|c|}
\hline
Potential & Mean Number of \efolds  & Probability of large field exploration  \\ \hline \hline
\multirow{2}{*}{Plateau 
}          &       \multirow{2}{*}{always infinite}      &  $0$ if $D\leq 2$, finite if $D>2$            \\ 
& & non-negligible if $D\gtrsim 2+\mathcal{O}(0.1)/ v_{\infty}$ \\
\hline
\multirow{2}{*}{Monomial $v\propto r^p$}    &      finite if $D<p$   &  $0$ if $D\leq 2$, finite if $D>2$  \\
 & $\ $infinite if $D\geq p$ & non-negligible if $D\gtrsim 1+p/ v_\uin$
\\ \hline
Hilltop          &     always finite    & \cellcolor{grey!25} \\ \hline
\end{tabular}  
\caption{Mean number of \efolds realised in $v(r)$ potentials and probability of exploring arbitrarily large-field regions of the potential, when $D$ fields are present.}   
\label{table:v(r):Nmean}         
\end{table}
\end{center}
The validity of the classical limit relies on the assumption that the second integral in \Eq{eq:v(r):sol} is dominated by its contribution close to the lower bound $r^\prime$. If this is correct, this means that the upper bound, $r_+$, can be removed to infinity without affecting the leading order result, providing a well-defined regularisation procedure. In the left panel of \Fig{fig:Nmean:v(r)}, one can see that for $D=1$, the curves saturate to an asymptotic behaviour when $v_+$ increases, and such a procedure seems therefore to exist. However, for $D=7$, the result does not seem to converge when $v_+$ increases. This is why in the right panel of \Fig{fig:Nmean:v(r)}, the mean number of \efolds is displayed as a function of $v_+$ for quartic $v(r)$-potentials, for a fixed $v_\uin=1$ and a few values of $D$. One can see that when $D<4$, $\langle\mathcal{N}\rangle$ goes to a constant value when $v_+\rightarrow\infty$, while when $D\geq4$, it diverges. This confirms that the number of fields plays an important role in determining whether the limit $r_+\rightarrow\infty$ is finite or not.

More precisely, the mean number of \efolds is finite if the integrand of \Eq{eq:v(r):sol} is integrable when $r^{\prime\prime}\rightarrow\infty$, that is to say if $r^{D-1}/v(r)$ is an integrable function. This criterion depends on the number of fields $D$, as already noticed, but also on the large-field behaviour of the potential. Let us distinguish the three following cases:
\begin{itemize}
\item If, at large-field value, the potential is of the ``Plateau'' type and $v$ goes to a constant value $v_\infty>0$, then $r^{D-1}/v(r)$ is never integrable and an infinite mean number of \efolds is always realised, regardless of the number of fields. 
\item If, at large-field value, the potential is of the monomial type $v\propto r^p$, then $r^{D-1}/v(r)$ is integrable only when $D<p$, and an infinite mean number of \efolds is realised as soon as more than $p$ fields are present. This is consistent with the previous discussion about the right panel of \Fig{fig:Nmean:v(r)}.
\item If the potential is of the ``hilltop'' type and symmetric around $0$, as explained above, $r_+$ has to be replaced by $0$ in \Eq{eq:v(r):sol}. In this case, the integrability of $r^{D-1}/v(r)$ needs to be checked around $0$ instead of infinity. If $v$ is finite at $r=0$ this is always the case, hence the mean number of \efolds is never infinite in such potentials.
\end{itemize}
The situation is summarised in table~\ref{table:v(r):Nmean}. In a large class of potentials (plateau potentials and some monomial potentials), the mean number of \efolds is infinite, and we call this phenomenon ``infinite inflation''. Let us notice that this is different from ``eternal inflation''~\cite{Steinhardt:1982kg, Vilenkin:1983xq, Guth:1985ya, Linde:1986fc} where volume weighting is included and the diverging quantity is the physical volume of the inflating part of the Universe, not $\langle \mathcal{N} \rangle$. Infinite inflation implies eternal inflation but is a stronger statement. For example, eternal inflation can be realised in hilltop models~\cite{Vilenkin:1983xq, Barenboim:2016mmw} while, as we have just shown, infinite inflation never occurs in such potentials. 

Another important remark is that for monomial potentials, whether infinite inflation occurs crucially depends on the number of fields, which therefore plays the role of an ``order parameter'' (as illustrated below in \Fig{fig:pwall:v(r)}). The number of dimensions is a critical parameter for many stochastic processes (for instance in recurrence problems~\cite{Polya:1921}) and this may therefore not be so surprising. The key feature is that the more fields, the larger the volume in field space to realise inflation and the more common infinite inflation.

Beyond the physical implications related to the possibility of realising arbitrarily large number of \efolds\cite{Polarski:1992dq, Ringeval:2010hf, Enqvist:2012xn}, infinite inflation raises the issue of practical calculability of observables. Indeed, since the correlation functions of scalar adiabatic fluctuations are related to the moments of the number of $e$-folds, see \Sec{sec:intro}, it is not clear what the predictions for these observables are when those moments are infinite. How these infinities regularise or not is in fact a non-trivial question that we investigate separately in \Ref{Vennin:2016wnk}. At this stage however, let us notice that infinite inflation may suggest that the system explores regions of the potential that are far away from what its classical trajectory would allow it to reach, and that observables may be sensitive to the physics at play in these remote regions. For this reason, we now study how likely it is to explore large-field regimes in stochastic multiple field inflation.
\subsection{Large Field Exploration}
\label{sec:wallproba}
\begin{figure}[t]
\begin{center}
\includegraphics[width=0.49\textwidth]{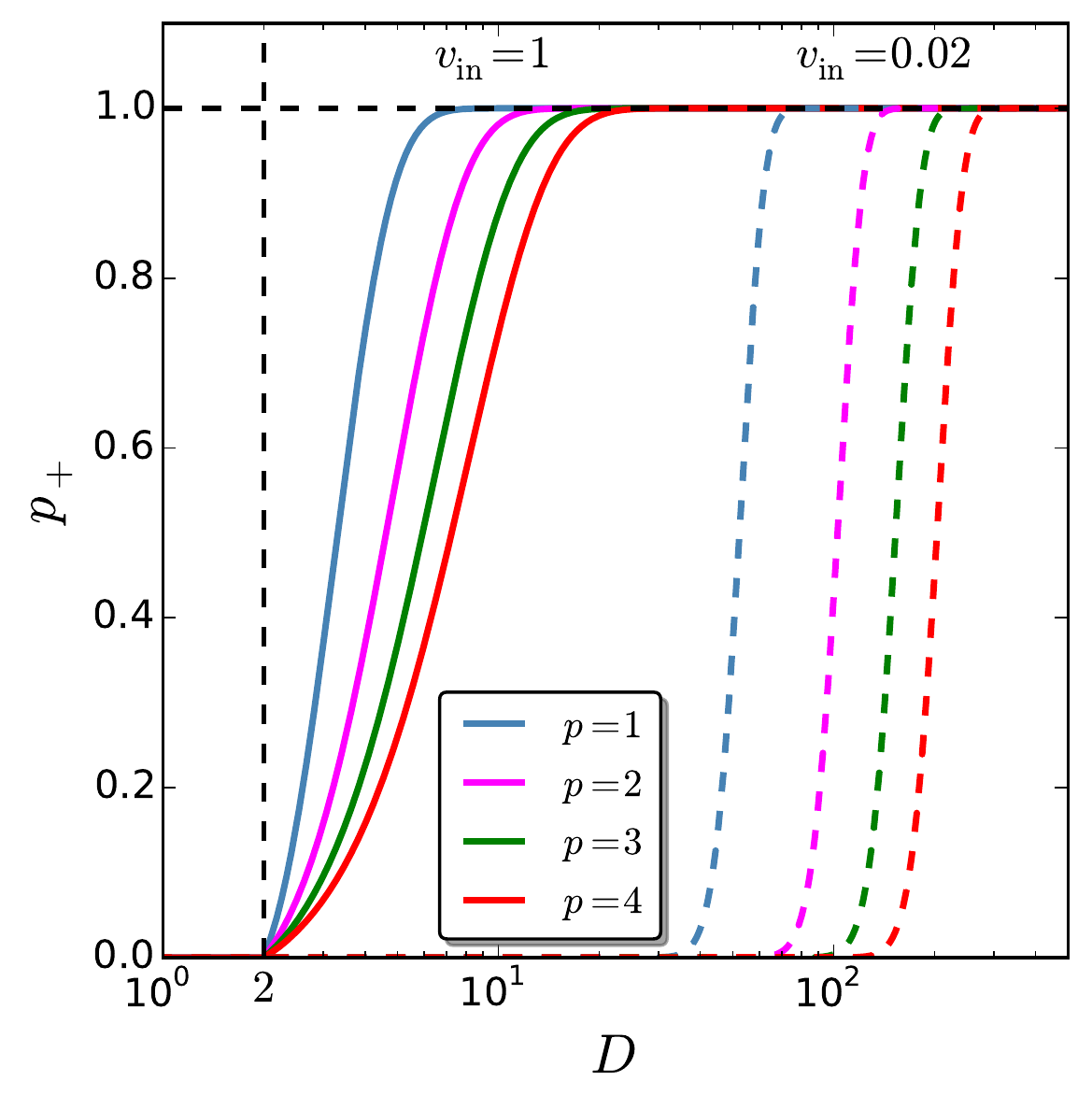}
\includegraphics[width=0.478\textwidth]{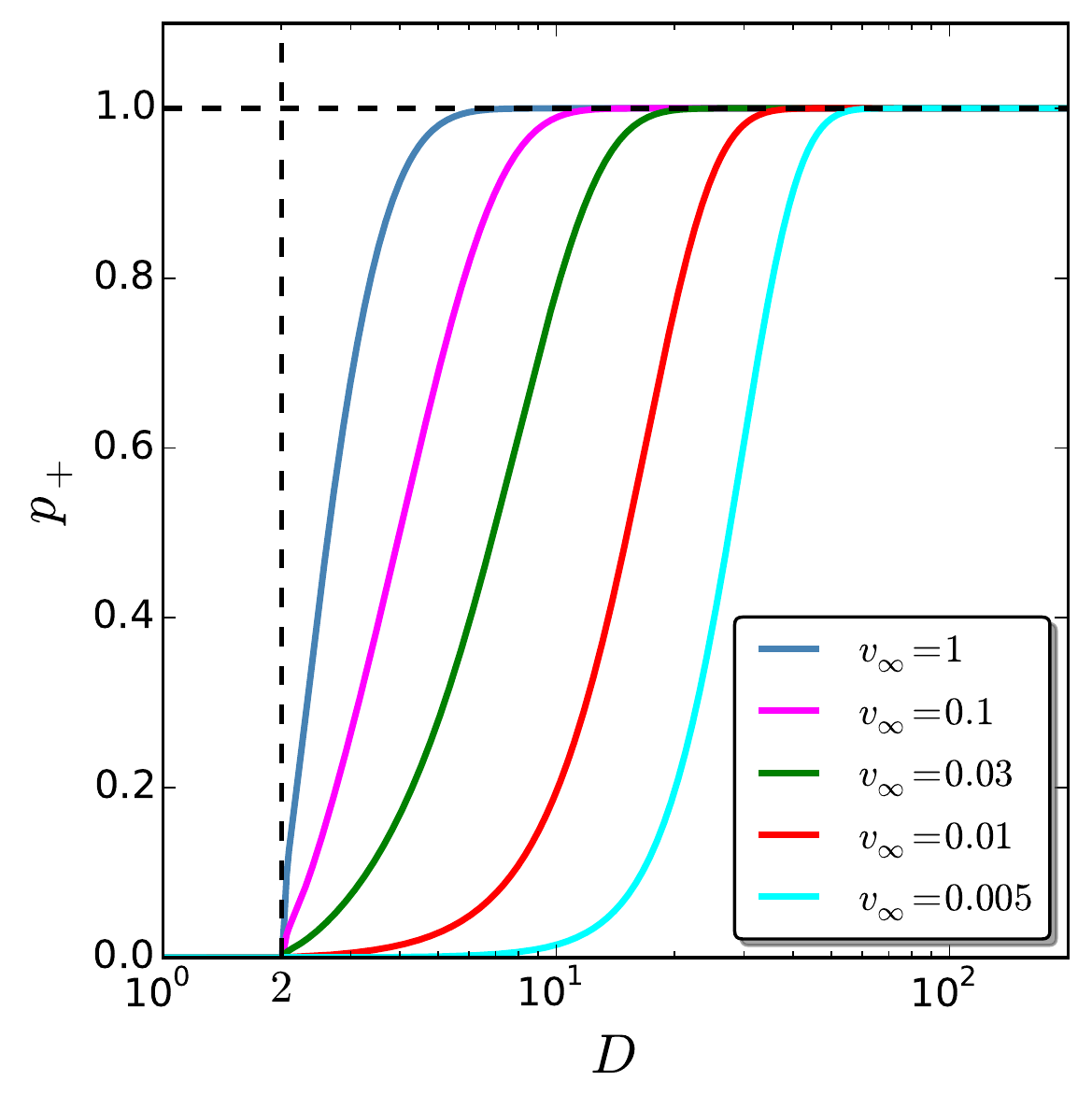}
\caption{Probability~(\ref{eq:v(r):pwall}) of exploring large-field regions of the potential $r_+$ when $r_+\rightarrow \infty$, as a function of the numbers of fields $D$. The left panel stands for monomial potentials $v(r)\propto r^p$, where a few values of $p$ are displayed. The first set of curves (solid lines) correspond to choosing the initial value of $r$ such that $v_\uin=1$, while $v_\uin=0.1$ in the second set of curves (dashed lines). The lower boundary condition is taken to be such that $r_-/\Mp=p/\sqrt{2}$ (end of inflation by slow-roll violation) but one can check that its precise value plays a negligible role. This probability is always non-zero when $D> 2$, but in practice, it is non-negligible only when $D\gtrsim 1+p/v_\uin$. The right panel stands for a plateau potential, the Starobinsky model, for which $v=v_{\infty}[1-\exp(-\sqrt{2/3}r/\Mp)]^2$.  The lower boundary condition is taken to be such that $r_-/\Mp=\sqrt{3/2}\ln(1+2/\sqrt{3})$ (end of inflation by slow-roll violation), and the initial value of $r$ is taken $50$ (classical) \efolds before the end of inflation. Several values of $v_\infty$ are displayed, and the probability $p_+$ is non-negligible only when $D\gtrsim 2+\mathcal{O}(0.1)/v_\infty$.}
\label{fig:pwall:v(r)}
\end{center}
\end{figure}
In this section, we derive the probability $p_+(r)$ that, starting from $r$, the system bounces at least once against the reflecting wall located at $r_+$ before exiting inflation at $r_-$ (or alternatively, if an absorbing wall is located at $r_+$, the probability that the system exits inflation at $r_+$ rather than $r_-$). In \Sec{sec:BoundaryConditions}, it was explained that $p_+$ is given by the solution of \Eq{eq:FPB:equadiff} with boundary conditions $p_+(r_-)=0$ and $p_+(r_+)=1$. Making use of \Eq{eq:Fpop:radial:v(r)}, one obtains
\begin{align}
\label{eq:v(r):pwall}
p_+\left(r\right)=\displaystyle\dfrac{\displaystyle\int_{r_-}^r {r^\prime}^{1-D}\ee^{-\frac{1}{v({r^\prime})}}\dd {r^\prime}}{\displaystyle\int_{r_-}^{r_+} {r^\prime}^{1-D}\ee^{-\frac{1}{v({r^\prime})}}\dd {r^\prime}}\, .
\end{align}
When the upper boundary condition $r_+$ is removed to infinity, one obtains a non-vanishing probability $p_+$ if the function $r^{1-D}$ is integrable (assuming that $v$ has a positive limit at infinity). Contrary to the case of infinite inflation in \Sec{sec:v(r):InfiniteInflation}, this condition is independent of the shape of the potential at large-field value, and $p_+>0$ as soon as strictly more than $2$ fields are present. This information is added in table~\ref{table:v(r):Nmean}, and in \Fig{fig:pwall:v(r)}, \Eq{eq:v(r):pwall} is displayed for monomial potentials $v\propto r^p$ (left panel, for different values of $p$ and $v_\uin$) and a plateau potential, the Starobinsky model~\cite{Starobinsky:1980te}, $v=v_\infty[1-\exp(-\sqrt{2/3}r/\Mp)]^2$ (right panel, for different values of $v_\infty$), when $r_+$ is removed to infinity. One can check that when $D\leq 2$, $p_+=0$. When $D>2$, strictly speaking, $p_+>0$, but one can see that $p_+$ is non-negligible only when $D$ is larger than some value that depends on the parameters of the potential and on the initial field \textit{vev}. Schematically, this value is realised when the integrand of the integrals in \Eq{eq:v(r):pwall} is maximal at $v_\uin$. In monomial potentials, this leads to the conclusion that $p_+$ is non-negligible when 
\begin{align}
D\gtrsim 1+\frac{p}{v_\uin}\, ,
\end{align}
while in plateau potentials, one obtains the condition
\begin{align}
D\gtrsim 2+\frac{\mathcal{O}(0.1)}{v_\infty}\, .
\end{align}
One can numerically check that, indeed, these expressions provide good estimates of the point where $p_+$ starts to be non-negligible. This shows that including more fields increases the probability to explore large-field regions of the potential, but for sub-Planckian energy scales, one needs a very large number of fields to obtain a substantial probability. For example, if one normalises the overall mass scale of the potentials to fit the measured amplitude of the scalar power spectrum~\cite{Ade:2015xua} and starts the evolution $50$ (classical) \efolds before the end of inflation, one finds $v_\uin\sim10^{-11}p$ for monomial models and $v_{\infty}\sim 10^{-12}$ for the Starobinsky potential, so that $10^{11}$ fields would be required to obtain appreciable values of $p_+$, a very large number indeed.

Of course, from a model building perspective, the shape of the potential may be very different at very large field outside the observational window than what cosmological observations constrain at smaller field values (for example~\cite{Broy:2014sia, Coone:2015fha}, the potential may be of the Plateau type where the scales probed in the CMB cross the Hubble radius, but of the monomial type at larger field), and if inflation starts high enough in the potential, large-field exploration, enhanced by the presence of multiple fields, may become likely. But the above results suggest that, in the simplest setups, cosmological observations at small (\ie sub-Planckian) energies carry limited information about the physics taking place at much higher energy (at least through stochastic effects). We further investigate this question in \Ref{Vennin:2016wnk}.
\section{Inhomogeneous End of Inflation}
\label{sec:InhomogeneousEnd}
In \Sec{sec:InfiniteInflation}, we have considered the case of $v(r)$ potentials where the dynamics is governed by the ``radial'' field $v$ only, and angular independent solutions can be found. In this section, we study situations where both $v$ and $\theta_j$ play a role, and study the simple setup where the inflationary dynamics is effectively driven by a single field $\phi$ while extra fields $\chi_j$ (to be identified with the ``angles'' $\theta_j$) only appear at the surface defining the end of inflation. This notably allows one to describe ``inhomogeneous end of inflation''~\cite{Dvali:2003ar} where inhomogeneities induced from the additional light fields on the end-surface make additional contributions to curvature perturbations on super-Hubble scales. 

In the terminology introduced in \Sec{sec:harmonic}, this case is called ``linear potential'' (\ie $v$ depends on a linear combination of the fields only) and is described in \Sec{sec:linear}. In \Eq{eq:Fpop:linear}, one can see that the ``angular'' sector is only affected by a pure diffusion term, which suggests that some insight may be gained by Fourier transforming the functions
\begin{align}
f_n\left(\phi, \bm{\chi} \right)&=\int \dd^{D-1} \bm{k} \ee^{-i  \bm{k}\cdot\bm{\chi}} f_n^{\bm{k} }(\phi)\, .
\label{eq:linearpot:FourierExpansion}
\end{align}
Plugging this ansatz into \Eq{eq:Fpop:linear}, \Eq{eq:PDE} gives rise to the set of recursive ordinary differential equations
\begin{align}
v \left(f_n^{\bm{k}}\right)^{\prime\prime}(\phi)-\frac{v^\prime(\phi)}{v(\phi)} \left(f_n^{\bm{k}}\right)^{\prime}(\phi) - k^2 v f_n^{\bm{k}}(\phi)=-\frac{n}{\Mp^2}f_{n-1}^{\bm{k}}(\phi)\, ,
\label{eq:linearpot:ode}
\end{align}
where $k^2\equiv \left\vert \bm{k}\right\vert^2$. If the end-surface $\partial\Omega_-$ is parametrised by the function $\phi_-(\bm{\chi})$, and if the inflationary domain is limited from above at the reflecting surface $\partial\Omega_+$ defined by $\phi=\phi_+$, these equations need to be solved with the boundary conditions 
\begin{align}
f_n\left [\phi_-\left(\bm{\chi}\right),\bm{\chi}\right]=0\, ,\quad\quad \frac{\partial f_n}{\partial\phi}\left(\phi_+,\bm{\chi}\right)=0\, .
\label{eq:linearpot:boundarycondition}
\end{align}
The procedure one needs to carry out is therefore the following: solve \Eq{eq:linearpot:ode} for all $\bm{k}$s in terms of two integration constants each, plug the solutions into \Eq{eq:linearpot:FourierExpansion}, and use \Eq{eq:linearpot:boundarycondition} to set all integration constants simultaneously. In practice, such a calculation may need to partly rely on numerical analysis, but it is still more straightforward (and numerically less expensive) than having to solve the full partial differential equations~(\ref{eq:PDE}).
\subsection{The Example of Exponential Potentials}
\begin{figure}[t]
\begin{center}
\includegraphics[width=0.48\textwidth]{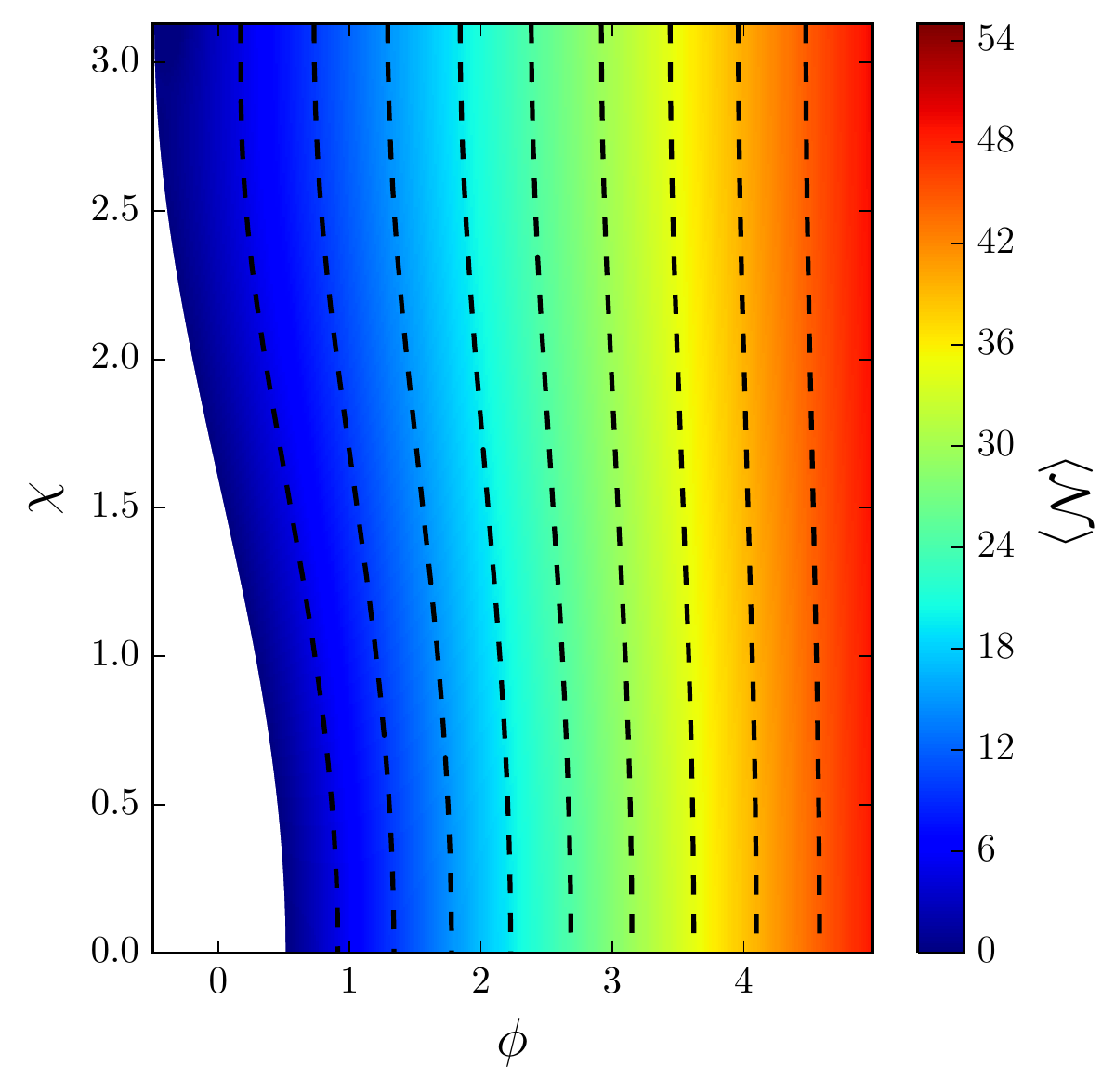}
\includegraphics[width=0.48\textwidth]{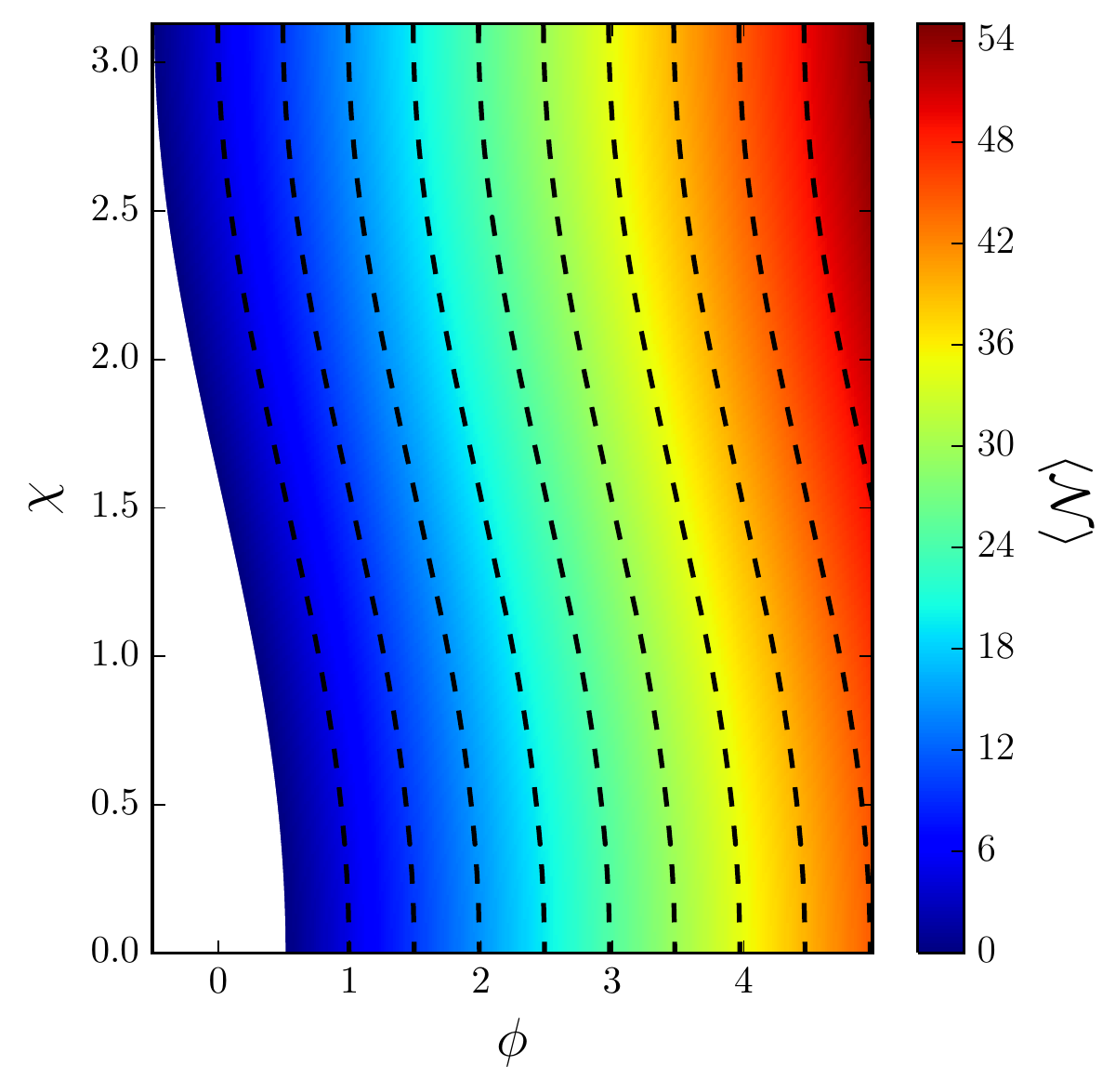}
\caption{Mean number of \efolds  $\langle\mathcal{N}\rangle$ realised in the single-field ``power-law'' potential $v=v_\uend \ee^{\alpha\phi}$, when the end of inflation is modulated by an extra field $\chi$ through the function $\phi_-(\chi)=\mu\cos(\chi/\chi_0)$. The left panel corresponds to the full stochastic result~(\ref{eq:exponential:sol}), where the integration constants $A_0$, $A_k$, $B_0$, $B_k$ are obtained imposing \Eqs{eq:linearpot:boundarycondition}. The right panel corresponds to the classical limit~(\ref{eq:modulated:classicalLimit}), which provides a good approximation to the stochastic result when $v\ll 1$. The parameter values in both panels are $\alpha=0.1$, $\chi_0/\Mp=1$, $\mu/\Mp=0.5$ and $v_\uend=0.05$. The black dashed lines correspond to various level lines of $\langle\mathcal{N}\rangle$, and help to better compare the two results. In particular, when $v$ increases, one can see that the dependence on the initial value of $\chi$ gets smeared out by the stochastic effects.}
\label{fig:modulatedReheating}
\end{center}
\end{figure}
In order to illustrate how the above procedure works in practice, let us consider the case of ``power-law inflation''~\cite{Abbott:1984fp} where the potential is of the exponential type $v\propto\ee^{\alpha\phi/\Mp}$. To be explicit, we study the situation where one extra field $\chi$ modulates the end-surface through 
\begin{align}
\phi_-\left(\chi\right) = \phi_\uend +\mu \cos\left(\frac{\chi}{\chi_0}\right)\, ,
\end{align} 
where $\mu$ is a modulation parameter (when $\mu\rightarrow 0$, one recovers the standard single-field setup), and $\chi_0$ is the scale over which the modulation takes place. In this case, solutions of  \Eq{eq:linearpot:ode} that are $2\pi\chi_0$-periodic in $\chi$ can be found, and one can replace the continuous Fourier transform of \Eq{eq:linearpot:FourierExpansion} by a discrete Fourier sum over integer numbers $k$, $f_n(\phi,\chi)=\sum_k\ee^{-ik\chi/\chi_0} f_n^k(\phi)$. Let us also note that since the exponential potential is conformally invariant, shift symmetry in the inflaton field value allows us to take $\phi_\uend=0$ without loss of generality, and write $v=v_\uend\ee^{\alpha\phi/\Mp}$. For the mean number of \efolds $f_1$, recalling that $f_0=1$, \Eq{eq:linearpot:ode} then gives rise to
\begin{align}
 \left(f_1^{\bm{k}}\right)^{\prime\prime}(\phi)-\frac{\alpha}{\Mp v\left(\phi\right)}\left(f_1^{\bm{k}}\right)^{\prime}(\phi) - \frac{k^2}{\chi_0^2} f_1^{\bm{k}}(\phi)=-\frac{\delta_{k,0}}{\Mp^2 v\left(\phi\right)}\, .
\label{eq:linearpot:ode:exppot}
\end{align}
When $k=0$, the solution can be expressed in terms of the exponential integral function $\Ei$, while when $k\neq 0$, the solution is given in terms of confluent hypergeometric functions ${}_1F_1$. Requiring that $f_1$ is real, one obtains
\begin{align}
f_1(\phi,\chi)=&A_0+\frac{\phi}{\alpha\Mp}+B_0\Ei\left[-\frac{1}{v(\phi)}\right]
\nonumber\\ &+
\sum_{k=1}^\infty \left\lbrace A_k v^{\frac{k \Mp}{\alpha\chi_0}}(\phi)\, {}_1F_1\left[-\frac{k \Mp}{\alpha\chi_0},1-2\frac{k \Mp}{\alpha\chi_0},-\frac{1}{v\left(\phi\right)}\right]
\right.\nonumber \\ & \left.
+B_k v^{-\frac{k \Mp}{\alpha\chi_0}}(\phi)\, {}_1F_1\left[\frac{k \Mp}{\alpha\chi_0},1+2\frac{k \Mp}{\alpha\chi_0},-\frac{1}{v\left(\phi\right)}\right]
\right\rbrace\cos\left({k\frac{\chi}{\chi_0}}\right)\, ,
\label{eq:exponential:sol}
\end{align}
where $A_0$, $A_k$, $B_0$, $B_k$ are integration constants, that must be fixed making use of \Eqs{eq:linearpot:boundarycondition}. At this stage, one has to proceed numerically. In practice, if the summation over $k$ in \Eq{eq:exponential:sol} is truncated at order $k_\umax$, one has $2(k_\umax+1)$ integration constants to fix. One can choose $k_\umax+1$ values of $\chi$ uniformly distributed in $[0,\pi\chi_0]$, and evaluate both parts of \Eqs{eq:linearpot:boundarycondition} at these values. This gives rise to $2(k_\umax+1)$ linear equations for the $2(k_\umax+1)$ integration constants, that one can solve with standard matrix inversion methods. One then increases $k_\umax$ until a sufficient level of convergence is reached.

The result of such a procedure is displayed in the left panel of \Fig{fig:modulatedReheating}, where the mean number of \efolds is plotted as a function of the initial values of $\phi$ and $\chi$, with $\alpha=0.1$, $\chi_0/\Mp=1$, $\mu/\Mp=0.5$ and $v_\uend=0.05$ (this last value does not lead to the right scalar power spectrum amplitude~\cite{Ade:2015xua}, but it is used to make clearer the effects we want to comment on). The value of $\phi_+$ has been taken to be sufficiently large so that it does not play any role, which is possible since the model is effectively single-field and non-plateau during inflation, as follows from the discussion in \Sec{sec:InfiniteInflation}. One has taken $k_\umax=100$, but the result is already very well converged when $k_\umax \gtrsim 8$. The black dashed lines are various level lines of $\langle\mathcal{N}\rangle$ and have been superimposed to guide the eye. The white region at the bottom left corresponds to $\phi<\phi_-(\chi)$, which lies outside of the inflationary domain. One may also note that only the region $0\leq\chi\leq\pi\chi_0$ is displayed, since the result for other values of $\chi$ can easily be inferred using the symmetry and periodicity properties of $f_1$. 
\subsection{Smearing Out the Modulating Field}
In the left panel of \Fig{fig:modulatedReheating}, one can notice that, going from the left to the right, the level lines of $\langle\mathcal{N}\rangle$ at first follow the modulation of the end-surface, and then become more and more straight. This result can be understood in terms of the two limits $v\ll 1$ and $v\gg 1$.

In the classical limit $v\ll 1$, the diffusion term acting on the extra field $\chi$ in \Eq{eq:Fpop:linear} is negligible, and $\chi$ freezes to its initial value. In this limit, the point where the system exits inflation in field space becomes deterministic and is simply given by $\phi_-(\chi_\uin)$. A similar expression to \Eq{v(r):Ncl} can therefore be obtained, except that the lower bound now explicitly depends on $\chi$,
\begin{align}
N_\ucl = \frac{1}{\Mp^2}\int_{\phi_-(\chi)}^\phi\frac{v\left(\phi^\prime\right)}{v^\prime\left(\phi^\prime\right)}\dd\phi^\prime\, .
\label{eq:modulated:classicalLimit}
\end{align}
For the power-law model under consideration, this gives rise to $N_\ucl=[\phi-\mu\cos(\chi/\chi_0)]/(\Mp\alpha)$. This quantity is displayed in the right panel of \Fig{fig:modulatedReheating} where one can check that, at small field where $v$ is not too large, it provides a good approximation to the full stochastic result given in the left panel.

In the stochastic dominated limit where $v\gg 1$, the diffusion term acting on $\chi$ becomes large, and quickly randomises the \textit{vev} of this extra field. In this regime, memory of the initial conditions on $\chi$ is quickly erased, which explains why the result becomes dependent on $\phi$ only and the level lines of $\langle \mathcal{N} \rangle$ in the left panel of \Fig{fig:modulatedReheating} tend to being merely vertical. Technically, one can check that the first line in \Eq{eq:exponential:sol}, which is the $0^\mathrm{th}$ (\ie $\chi$-independent) mode, provides the dominant contribution in the limit $v\gg 1$. This term exactly matches the one obtained in \Eq{eq:fn:sol:onefield} in a purely single-field setup. 

Therefore, stochastic effects tend to erase the presence of modulating fields by blurring their initial values and averaging the exit point over the end-surface. In practice, the size of the effect depends on how the scale over which modulation takes place (denoted $\chi_0$ in the present model) compares to the dispersion acquired by the modulating field at the end of inflation, but not so much on the size of the modulation itself (here denoted $\mu$). In the simple toy model discussed in this section, if one sets $v\simeq 10^{-10}$, corresponding to the value that would fit the measured scalar power spectrum amplitude, one finds\footnote{Here, the dispersion acquired by the freely diffusing $\chi$ field is given by $\sqrt{50}H/2\pi\simeq \sqrt{100 v}\Mp$, where $50$ is taken to be the number of \efolds between Hubble exit time of the scales probed in the CMB and the end of inflation, and $H/2\pi$ is the amplitude of the noise term in \Eq{eq:Langevin} which we assume to be roughly constant over these last $50$ \efolds of inflation.} that the effect is large if $\chi_0/\Mp\lesssim 10^{-4}$. The situation is therefore different from the purely single-field case~\cite{Vennin:2015hra} where, in the simplest setups, the stochastic effects must be subdominant since $v$ is observationally constrained to be small. Here, large stochastic corrections can be obtained even if $v\ll 1$, depending on the scales of the features in the end-surface.
\section{Conclusion}
\label{sec:conclusion}
Let us now summarise our main results. In this paper, we have investigated the inclusion of multiple fields in the stochastic inflationary framework. In particular, we have derived a hierarchy of diffusion equations that can be used to calculate the moments of the numbers of $e$-folds. This sets up the formalism that will be required to compute correlation functions of primordial curvature perturbations in the stochastic $\delta N$ approach in multi-field inflation, which we will investigate in \Ref{Vennin:2016wnk}. Compared to the single-field case, two main differences have been noted.

Firstly, the volume available in field space is strongly dependent on the number of fields $D$. For instance, for a given mass scale $\mu$, the volume associated to the region $\vert\bm{\phi}\vert<\mu$ is given by $\pi^{D/2}\mu^D/\Gamma(D/2+1)$. This is why the dimension often plays the role of a critical parameter, or an order parameter, in stochastic processes. This is for example the case in the well-known recurrence problems where it can be shown that, for a random walk on a $D$-dimensional lattice, the probability to return to the origin is $1$ for $D=1$ or $D=2$, and strictly smaller than $1$ for $D>2$ (and is then given by the so-called P\'olya's random walk constants~\cite{Polya:1921}). Thus $D=2$ can be seen as a critical dimension for the recurrence problems. A similar effect takes place in stochastic inflation, where we have shown that the probability to explore arbitrarily large field regions of the potential vanishes for $D\leq 2$ and is strictly non-zero when $D>2$, regardless of the potential. In practice, depending on the large-field profile of the potential, this probability is often much smaller than one (although not strictly vanishing) for initial sub-Planckian field values, even when a large number of fields is present. However, this illustrates why the number of fields plays a critical role in stochastic inflation and why including more fields tends to make the system explore regions of the potential that would otherwise be inaccessible with a single field. This notably translates into ``infinite inflation'', \ie the fact that the mean number of realised \efolds is infinite in all plateau models, and in monomial potentials $v\propto \phi^p$ if $p$ or more fields are present. Infinite inflation has important consequences for problems related to the total duration of inflation (which can be a crucial parameter in deriving the distribution of curvaton-type fields~\cite{Enqvist:2012xn, Vennin:2015egh} or other light spectator fields~\cite{Ringeval:2010hf} at the end of inflation, but also to determine whether pre-inflationary imprints on large scales can arise in ``just enough inflation'' types of scenarios~\cite{Polarski:1992dq}). Since the correlation functions of cosmological perturbations are related to the moments of \efolds number, this also means that infinite quantities appear in cosmological observables. Whether these infinities can be regularised and whether one needs to modify the large-field sector of the theory to make the results finite is therefore an important, number of fields dependent, question, that we address separately in a companion paper~\cite{Vennin:2016wnk}.

Secondly, including more than one scalar field offers more directions (than the classical one, aligned with the gradient of the potential in field space) along which the system can fluctuate. If quantum diffusion spreads the fields distribution along these extra directions on a distance (in field space) over which the potential does not vary much, the associated effect remains small. However, in the opposite case, large stochastic corrections can be produced. This situation was exemplified in \Sec{sec:InhomogeneousEnd} in the case where inflation ends inhomogeneously, \ie for a value of the inflaton field $\phi$ that depends on a modulating field $\chi$. If the dispersion in the $\chi$ field direction at the end of inflation is larger than the $\chi$ \textit{vev} scale over which the surface of end of inflation is modulated, large stochastic effects are obtained. Let us stress that this can happen even at sub-Planckian energies, contrary to what is found in purely single-field setups. In \Ref{Vennin:2015hra} indeed, it was shown that, even if stochastic effects can shift the location of the observational window along the inflationary potential, once the observational window is fixed, single-field stochastic corrections scale with $v$, which is aways a very small quantity. In multiple field scenarios, we have found that this property does not hold, and that stochastic corrections to cosmological observables can a priori be large within the observational window~\cite{Vennin:2016wnk}. This opens interesting possibilities for probing quantum effects on inflationary dynamics.
\acknowledgments
M.N.\ acknowledges financial support from the research council of University of Tehran. H.A., V.V. and D.W. acknowledge financial support from STFC grants ST/K00090X/1 and ST/N000668/1.
\appendix
\section{First Passage Time from Fokker-Planck Equation}
\label{app:FPT}
In this section, \Eqs{eq:PDE} are derived making use of ``first passage time analysis'' techniques. We consider the situation depicted in \Fig{fig:sketchFirstPassageTime}, for a stochastic process described by \Eq{eq:Langevin}, or equivalently, \Eq{eq:FokkerPlanck}.

The first step is to consider the transition or conditional probability $\rho(\phi_i,N\vert \phi_i^\uin,N_\uin)$ that the system is at $\phi_i$ at time $N$ given that it started off at $\phi_i^\uin$ at time $N_\uin$. Since \Eq{eq:Langevin} describes a Markovian process, this quantity is only a function of $N-N_\uin$. The Fokker-Planck equation~(\ref{eq:FokkerPlanck}), valid for any initial condition, gives, in this specific case, 
\begin{align}
\frac{\partial}{\partial N}\rho\left(\phi_i,N\vert \phi_i^\uin,N_\uin\right)=\mathcal{L}_\mathrm{FP}\left(\phi_i\right)\cdot\rho\left(\phi_i,N\vert \phi_i^\uin,N_\uin\right)\, . 
\label{eq:FP:rho}
\end{align}
We now want to write a similar equation but where the time derivative acts on the first time argument, $N_\uin$. Let us start from the Chapman-Kolmogorov relation
\begin{align} 
 \rho(\phi_i,N\vert \phi_i^\uin,N_\uin) = \int \dd\bar{\phi}_i  \rho(\phi_i,N\vert \bar{\phi}_i,\bar{N})  \rho(\bar{\phi}_i,\bar{N}\vert \phi_i^\uin,N_\uin)
\end{align} 
which simply states that any process starting at $\phi_i^\uin$ at $N_\uin$ and ending up at $\phi_i$ at $N$ goes through some point $\bar{\phi}_i$ at some intermediate time $\bar{N}$, and we integrate over all intermediary points $\bar{\phi}_i$. When one differentiates this relation with respect to $\bar{N}$, the left hand side vanishes, and one obtains
\begin{align}
 0 &= \int \dd\bar{\phi}_i \left[ \frac{\partial \rho(\phi_i,N\vert \bar{\phi}_i,\bar{N})}{\partial \bar{N}}  \rho(\bar{\phi}_i,\bar{N}\vert \phi_i^\uin,N_\uin)
 + \rho(\phi_i,N\vert \bar{\phi}_i,\bar{N}) \frac{\partial \rho(\bar{\phi}_i,\bar{N}\vert \phi_i^\uin,N_\uin)}{\partial \bar{N}} \right]
\\ 
&= \int \dd\bar{\phi}_i \left[ \frac{\partial \rho(\phi_i,N\vert \bar{\phi}_i,\bar{N})}{\partial \bar{N}}  \rho(\bar{\phi}_i,\bar{N}\vert \phi_i^\uin,N_\uin)
 + \rho(\phi_i,N\vert \bar{\phi}_i,\bar{N}) \mathcal{L}_\mathrm{FP}(\bar{\phi}_i)\cdot \rho(\bar{\phi}_i,\bar{N}\vert \phi_i^\uin,N_\uin)\right] 
 \, ,
 \label{eq:ChapmannKolmogorov:FP}
\end{align}
where, in the second line, we have used the Fokker-Planck equation~(\ref{eq:FP:rho}). The second term in the integral of \Eq{eq:ChapmannKolmogorov:FP} can be integrated by parts making use of the adjoint Fokker-Planck operator $\mathcal{L}_\mathrm{FP}^\dagger$ defined as
\begin{align}
\label{eq:FPoperator:adjoint}
\frac{1}{\Mp^2}\mathcal{L}_\mathrm{FP}^\dagger(\phi_i) = -\sum_i\frac{v_{\phi_i}}{v}\frac{\partial}{\partial\phi_i}+v\sum_i\frac{\partial^2}{\partial\phi_i^2}\, ,
\end{align}
and one obtains the adjoint Fokker-Planck equation
\begin{align}
\frac{\partial}{\partial N_\uin}\rho(\phi_i,N\vert \phi_i^\uin,N_\uin) = -\mathcal{L}^\dagger_{\mathrm{FP}}\left(\phi_i^\uin\right)\cdot \rho(\phi_i,N\vert \phi_i^\uin,N_\uin)\, .
\label{eq:FP:adjoint}
\end{align}

The next step is to introduce the survival probability $S(N)$ of having not yet crossed $\partial\Omega$ at time $N$,
\begin{align}
S(N)=\int_{\Omega} \rho(\phi_i,N\vert \phi_i^\uin,N_\uin)\dd\phi_i\, .
\end{align}
This corresponds to the probability of having $\mathcal{N}>N$. If $P(N)$ denotes the probability distribution associated with $\mathcal{N}$, this means that 
\begin{align}
S(N)=\int_N^\infty P(N^\prime)\dd N^\prime\, .
\end{align}
By differentiating this expression with respect to $N$, one obtains
\begin{align}
P(N) = -\frac{\dd}{\dd N}S(N) =- \int_{\Omega} \frac{\partial}{\partial N} \rho(\phi_i,N\vert \phi_i^\uin,N_\uin)\dd\phi_i\, .
\end{align}
The $n^\mathrm{th}$ moment of $\mathcal{N}$ can therefore be expressed as
\begin{align}
\label{eq:Nn:1}
\left\langle \mathcal{N}^n \right\rangle (\phi_i^\uin) &=
\displaystyle \int_{N_\uin}^\infty N^n P(N) \dd N
= - \int_{N_\uin}^\infty N^n \dd N \int_{\Omega} \dd\phi_i\frac{\partial}{\partial N} \rho(\phi_i,N\vert \phi_i^\uin,N_\uin) 
\\  
& =  n \displaystyle \int_{N_\uin}^\infty N^{n-1} \dd N \int_{\Omega} \dd\phi_i \rho(\phi_i,N\vert \phi_i^\uin,N_\uin) \, ,
\end{align}
where integration by parts has been performed, with the requirement that $\rho$ vanishes on  $\partial\Omega$ (absorbing boundary conditions). By applying the adjoint Fokker-Planck operator $\mathcal{L}_\mathrm{FP}^\dagger(\phi_i^\uin)$ on this relation and making use of the adjoint Fokker-Planck equation~(\ref{eq:FP:adjoint}), one obtains
\begin{align}
\mathcal{L}_\mathrm{FP}^\dagger(\phi_i^\uin)\cdot\left\langle \mathcal{N}^n \right\rangle (\phi_i^\uin)  & =-
n \int_{N_\uin}^\infty N^{n-1} \dd N \int_{\Omega} \dd\phi_i \frac{\partial}{\partial N_\uin} \rho(\phi_i,N\vert \phi_i^\uin,N_\uin)\, .
\end{align}
At this stage, let us recall that one is dealing with a Markovian process, for which the transition probability depends on $N-N_\uin$ only. One then has
\begin{align}
\label{eq:FPT:Markov}
\mathcal{L}_\mathrm{FP}^\dagger(\phi_i^\uin)\cdot\left\langle \mathcal{N}^n \right\rangle (\phi_i^\uin)  & = -
n \int_{N_\uin}^\infty N^{n-1} \dd N \int_{\Omega} \dd\phi_i \frac{\partial}{\partial N_\uin} \rho(\phi_i,N-N_\uin\vert \phi_i^\uin,0) \\
& = 
n \int_{N_\uin}^\infty N^{n-1} \dd N \int_{\Omega} \dd\phi_i \frac{\partial}{\partial N} \rho(\phi_i,N-N_\uin\vert \phi_i^\uin,0)
\\ & 
 = n \int_{N_\uin}^\infty N^{n-1} \dd N \int_{\Omega} \dd\phi_i \frac{\partial}{\partial N} \rho(\phi_i,N\vert \phi_i^\uin,N_\uin) \, .
 \label{eq:Nn:2}
\end{align}

When $n=1$, this gives rise to
\begin{align}
\mathcal{L}_\mathrm{FP}^\dagger(\phi_i^\uin)\cdot\left\langle \mathcal{N} \right\rangle (\phi_i^\uin)  & = \int_{N_\uin}^\infty \dd N \int_{\Omega} \dd\phi_i \frac{\partial}{\partial N} \rho(\phi_i,N\vert \phi_i^\uin,N_\uin) 
\\ & 
=\int_{\Omega} \dd\phi_i \left[\rho(\phi_i,\infty\vert \phi_i^\uin,N_\uin)-\rho(\phi_i,N_\uin\vert \phi_i^\uin,N_\uin) \right]
\\ & 
=\int_{\Omega} \dd\phi_i \left[0-\delta^D(\phi_i-\phi_i^\uin) \right]
= -1\, .
\label{FPT:neq1}
\end{align}
Here, we have used the fact that, in the infinite future, all realisations have crossed $\partial\Omega$,\footnote{This condition is necessary for the distribution $P(N)$ to be normalisable.} so that $\rho(\phi_i,\infty\vert \phi_i^\uin,N_\uin)=0$. When $n\geq 2$, from \Eq{eq:Nn:1}, one notices that $\langle \mathcal{N}^{n-1}\rangle$ appears in the right hand side of \Eq{eq:Nn:2}, giving rise to 
\begin{align}
\mathcal{L}_\mathrm{FP}^\dagger(\phi_i^\uin)\cdot\left\langle \mathcal{N}^n \right\rangle (\phi_i^\uin) = -n \left\langle \mathcal{N}^{n-1} \right\rangle (\phi_i^\uin)\, .
\end{align}
Defining 
\begin{align}
\label{eq:fn:meanNn}
f_n(\phi_i)\equiv \langle \mathcal{N}^n\rangle(\phi_i^\uin=\phi_i)\, ,
\end{align}
and recalling that the adjoint Fokker-Planck operator is given by \Eq{eq:FPoperator:adjoint}, the previous analysis shows that the $f_n$ functions satisfy the hierarchy of partial differential equations
\begin{align}
\sum_i\left(v\frac{\partial^2}{\partial\phi_i^2}-
\frac{v_{\phi_i}}{v}\frac{\partial}{\partial\phi_i}\right)f_n=-n\frac{f_{n-1}}{\Mp^2}
\label{eq:PDE:appendix}
\end{align}
for $n\geq 1$, where $f_0=1$.
\section{First Passage Time from Langevin Equation}
\label{app:FPT:Langevin}
The same results as the ones derived in \App{app:FPT} can also be obtained starting from the Langevin equation~(\ref{eq:Langevin}). In this section, we quickly sketch such a derivation for illustrative purpose, and to make comparison with \Ref{Vennin:2015hra} easier. 

Let $f(\bm{\phi})$ be a generic function of the fields value $\bm{\phi}=(\phi_1,\phi_2,\cdots,\phi_D)$. If $\bm{\phi}$ is a realisation of the stochastic process~(\ref{eq:Langevin}), its variation is given by
\begin{align}
\dd f(\bm{\phi}) &= \sum_i f_{\phi_i}\dd\phi_i +\frac{1}{2}\sum_{i,j}f_{\phi_i,\phi_j}\dd\phi_i\dd\phi_j+\order{\dd\bm{\phi}^3}\\
&= \Mp \sqrt{2v} \sum_i f_{\phi_i}\xi_i \dd N
-\Mp^2\sum_i f_{\phi_i}\frac{v_{\phi_i}}{v}\dd N
+\Mp^2 \sum_i v f_{\phi_i,\phi_i}\dd N +\order{\dd N^2}\, .
\end{align}
Integrating this relation between $N=0$ where $\bm{\phi}=\bm{\phi}_\uin$ and $N=\mathcal{N}$ where $\bm{\phi}=\bm{\phi}_\uend\in\partial\Omega_-$, one obtains the It\^o's lemma~\cite{ito1944}
\begin{align}
f\left(\bm{\phi}_\uend\in\partial\Omega_-\right) - f\left(\bm{\phi}_\uin\right) & = \int_0^\mathcal{N} \Mp \sqrt{2v} \sum_i f_{\phi_i}\xi_i \dd N
 + \int_0^\mathcal{N} \Mp^2 \left( v f_{\phi_i,\phi_i}-\sum_i f_{\phi_i}\frac{v_{\phi_i}}{v}\right)\dd N\, .
\label{eq:Ito}
\end{align}
Let us now apply this lemma to $f_1$, defined in \Sec{sec:FPT:moments} as being the solution of \Eq{eq:PDE} which vanishes along $\partial\Omega_-$. By definition, the first term in the left hand side of \Eq{eq:Ito} vanishes, and the integrand of the second integral of the right hand side is $-1$. This gives rise to
\begin{align}
\mathcal{N}=f_1\left(\bm{\phi}_\uin\right)+\int_0^\mathcal{N} \Mp \sqrt{2v} \sum_i f_{1,\phi_i}\xi_i \dd N\, .
\label{eq:Ito:f1}
\end{align}
By taking the stochastic average of this equation, one is led to 
\begin{align}
\mathcal{N}=f_1\left(\bm{\phi}_\uin\right)\, .
\end{align}
Note that the fact that the stochastic average of the integral term in \Eq{eq:Ito:f1} vanishes is not trivial a priori since not only the integrand but the upper bound of the integral itself is stochastic, but because the noises $\xi_i$ are uncorrelated at different times, this can be shown rigorously~\cite{Risken:1984book}. This demonstrates \Eq{eq:fn:meanNn} for $n=1$. 

Larger values of $n$ can be dealt with in a similar manner. Indeed, by squaring \Eq{eq:Ito:f1} and taking the stochastic average of it, one obtains
\begin{align}
\left\langle \mathcal{N}^2 \right\rangle=f_1^2\left(\bm{\phi}_\uin\right)+2\Mp^2 \left\langle \int_0^\mathcal{N} v \left(\sum_i f_{1,\phi_i}\right)^2 \dd N\right\rangle \, .
\label{eq:Ito:f2}
\end{align}
Let us now apply It\^o's lemma~(\ref{eq:Ito}) to $g_2\equiv f_2-f_1^2$, where $f_2$ is defined in \Sec{sec:FPT:moments} as being the solution of \Eq{eq:PDE} which vanishes along $\partial\Omega_-$. By definition, the first term in the left hand side of \Eq{eq:Ito} vanishes, and the integrand of the second integral of the right hand side is given by $-2 \sum_i f_{1,\phi_i}^2 v$. One obtains
\begin{align}
g_2\left(\bar{\phi}_\uin\right) & = 2\Mp^2\left\langle \int_0^\mathcal{N} v \left(\sum_i f_{1,\phi_i}\right)^2 \dd N\right\rangle
= \left\langle \mathcal{N}^2 \right\rangle - f_1^2\left(\bm{\phi_\uin}\right)\, ,
\end{align}
where \Eq{eq:Ito:f2} has been used in the last equality. Since $g_2=f_2-f_1^2$, this gives rise to $f_2(\bar{\phi}_\uin)= \langle\mathcal{N}^2 \rangle$, \ie \Eq{eq:fn:meanNn} for $n=2$. Applying the same method, one can iteratively proceed and extend the result to any $n$.
\section{First Passage Boundary}
\label{app:FPB}
In this section, we consider the case where $\partial\Omega$ is made of two (or more) disconnected pieces (say $\partial\Omega_-$ and $\partial\Omega_+$) and one wants to determine with which probability $p_+$ the system exits $\Omega$ crossing $\partial\Omega_+$ (or respectively, with which probability $p_-=1-p_+$ the system exits $\Omega$ crossing $\partial\Omega_-$).  

The same techniques as the ones employed in \App{app:FPT} can be employed to determine this quantity as a function of the initial conditions $\phi_\uin$ where the system starts off its evolution. By definition, $p_+$ corresponds to the probability that the system is somewhere along $\partial\Omega_+$ at time $N$, where $N$ is integrated over all possible values between $N_\uin$ and $\infty$,
\begin{align}
p_+(\phi_\uin)&=\int_{\phi_i\in\partial\Omega_+}\int_{N_\uin}^\infty\dd N\rho\left(\phi_i,N\vert \phi_i^\uin,N_\uin\right)\, .
\end{align}
Let us now apply the adjoint Fokker-Planck operator defined in \Eq{eq:FPoperator:adjoint} to this relation. Making use of \Eq{eq:FP:adjoint}, one obtains
\begin{align}
\mathcal{L}_{\mathrm{FP}}^\dagger\left(\phi_\uin\right) \cdot p_+(\phi_\uin)&=\int_{\phi_i\in\partial\Omega_+}\int_{N_\uin}^\infty\dd N \mathcal{L}_{\mathrm{FP}}^\dagger\left(\phi_\uin\right) \cdot \rho\left(\phi_i,N\vert \phi_i^\uin,N_\uin\right)\\
&=-\int_{\phi_i\in\partial\Omega_+}\int_{N_\uin}^\infty\dd N \frac{\partial}{\partial N_\uin} \rho\left(\phi_i,N\vert \phi_i^\uin,N_\uin\right)\\
&=-\int_{\phi_i\in\partial\Omega_+}\int_{N_\uin}^\infty\dd N \frac{\partial}{\partial N_\uin} \rho\left(\phi_i,N-N_\uin\vert \phi_i^\uin,0\right)\\
&=\int_{\phi_i\in\partial\Omega_+}\int_{N_\uin}^\infty\dd N \frac{\partial}{\partial N} \rho\left(\phi_i,N-N_\uin\vert \phi_i^\uin,0\right)\label{eq:FPB:markov}\\
&=\int_{\phi_i\in\partial\Omega_+}\int_{N_\uin}^\infty\dd N \frac{\partial}{\partial N} \rho\left(\phi_i,N\vert \phi_i^\uin,N_\uin\right)\\
&=\int_{\phi_i\in\partial\Omega_+} \left[ \rho\left(\phi_i,\infty\vert \phi_i^\uin,N_\uin\right) - \rho\left(\phi_i,{N_\uin}\vert \phi_i^\uin,N_\uin\right)\right]
=0\, .
\label{eq:FPB:0}
\end{align}
In \Eq{eq:FPB:markov}, similarly to what was performed in \Eq{eq:FPT:Markov}, one has used the fact that the stochastic process under consideration is Markovian, hence the transition probability depends on $N-N_\uin$ only. To obtain the final result~(\ref{eq:FPB:0}), one has also used the fact that, as mentioned below \Eq{FPT:neq1}, all realisations have crossed $\partial\Omega$ in the infinite future hence $\rho(\phi_i,\infty\vert \phi_i^\uin,N_\uin=0)$, together with the initial condition $\rho\left(\phi_i,{N_\uin}\vert \phi_i^\uin,N_\uin\right)=\delta^D(\phi_i-\phi_i^\uin)$ (and the assumption that $\phi_i^\uin \not\in \partial\Omega_+$, otherwise we already know that $p_+=1$ by definition). 

The probability $p_+(\phi_i)$ that the system first reaches $\partial\Omega_+$ starting from $\phi_i^\uin=\phi_i$ is therefore given by the solution of the ordinary differential equation
\begin{align}
\sum_i\left(v\frac{\partial^2}{\partial\phi_i^2}-
\frac{v_{\phi_i}}{v}\frac{\partial}{\partial\phi_i}\right)p_+=0\, ,
\label{eq:FPB:equadiff:app}
\end{align}
with boundary conditions $p_+=1$ on $\partial\Omega_+$ and $p_+=0$ on $\partial\Omega_-$.
\section{Spherical Coordinates in Arbitrary Dimension}
\label{app:spericalCoord}
If field space $\lbrace \phi_1,\phi_2,\cdots,\phi_D \rbrace$ has dimension $D$, the spherical coordinates $\lbrace r,\theta_1,\theta_2\cdots,\theta_{D-1}\rbrace $ are defined through
\begin{align}
\phi_1&=r\cos(\theta_1)\\
\phi_2 &= r\sin(\theta_1)\cos(\theta_2)\\
\phi_3 &= r\sin(\theta_1)\sin(\theta_2)\cos(\theta_3)\\
&\ \, \vdots\nonumber \\
\phi_{D-1}&=r\sin(\theta_1)\cdots\sin(\theta_{D-2})\cos(\theta_{D-1})\\
\phi_{D}&=r\sin(\theta_1)\cdots\sin(\theta_{D-2})\sin(\theta_{D-1})\, .
\end{align}
Here, $r\in[0,\infty[$, $\theta_j\in[0,\pi]$ for $1\leq j \leq D-2$ and $\theta_{D-1}\in [0,2\pi[$. The inverse transformation is given by
\begin{align}
r&=\sqrt{\sum_{i=1}^D\phi_i^2}\\
\theta_1&=\arccos\left(\frac{\phi_1}{r_1}\right)\\
\theta_2&=\arccos\left(\frac{\phi_2}{r_2}\right)\\
&\ \, \vdots\nonumber \\
\theta_{D-2}&=\arccos\left(\frac{\phi_{D-2}}{r_{D-2}}\right)\\
\theta_{D-1}&=
\begin{cases}
2\pi - \arccos\left(\frac{\phi_{D-1}}{r_{D-1}}\right)\quad\mathrm{if}\quad \phi_D< 0\ \mathrm{and}\ D>2\\
\arccos\left(\frac{\phi_{D-1}}{r_{D-1}}\right)\quad\mathrm{otherwise}\\
\end{cases}\, .
\end{align}
In these expressions, we have defined 
\begin{align}
r_j=\sqrt{\sum_{i=j}^D\phi_i^2}\, .
\end{align}
From here, the derivatives of the angular coordinates can be calculated, and one obtains
\begin{align}
\frac{\partial\theta_j}{\partial\phi_i} = 
\begin{cases}
0 &\mathrm{if}\ i<j\\
-\dfrac{r_{j+1}}{r_j^2} &\mathrm{if}\ i=j\\
\begin{cases}
-\dfrac{\phi_i\phi_j}{r_j^2r_{j+1}}\ \mathrm{if}\ j=D-1,\, D>2\, \mathrm{and}\, \phi_D<0\\
\dfrac{\phi_i\phi_j}{r_j^2r_{j+1}}\ \mathrm{otherwise}
\end{cases}
&\mathrm{if}\ i>j
\end{cases}\, .
\end{align}
The norm of the gradient of the angular coordinates appearing in \Eq{eq:Fpop:radial} is therefore given by
\begin{align}
\left\vert \bm{\nabla}(\theta_j)  \right\vert^2 = \frac{1}{r_j^2}
= \left[r\displaystyle\prod_{i=1}^{j-1}\sin(\theta_i)\right]^{-2}\, .
\end{align}
In particular, $\left\vert \bm{\nabla}(\theta_1)  \right\vert$ depends on $r$ only. In the same manner, the second derivatives of the angular coordinates can be calculated, from which their Laplacian,  appearing in \Eq{eq:Fpop:radial}, are found to be
\begin{align}
\Delta\theta_j=
\left(D-1-j\right)\frac{\phi_j}{r_j^2r_{j+1}}
= \frac{D-1-j}{r^2\tan\left(\theta_j\right)}\prod_{\ell=1}^{j-1}\sin^{-2}\left(\theta_\ell\right)\, .
\end{align}
In particular, this implies that $\Delta\theta_{D-1}=0$ and one recovers the fact that, when $D=2$, the only angular coordinate has vanishing Laplacian.
\bibliographystyle{JHEP}
\bibliography{formalism}

\providecommand{\href}[2]{#2}\begingroup\raggedright\begin{thebibliography}{10}

\bibitem{Starobinsky:1980te}
A.~A. Starobinsky, \emph{{A New Type of Isotropic Cosmological Models Without
  Singularity}},
  \href{http://dx.doi.org/10.1016/0370-2693(80)90670-X}{\emph{Phys. Lett.} {\bf
  B91} (1980) 99--102}.

\bibitem{Sato:1980yn}
K.~Sato, \emph{{First Order Phase Transition of a Vacuum and Expansion of the
  Universe}}, {\emph{Mon.Not.Roy.Astron.Soc.} {\bf 195} (1981) 467--479}.

\bibitem{Guth:1980zm}
A.~H. Guth, \emph{{The Inflationary Universe: A Possible Solution to the
  Horizon and Flatness Problems}},
  \href{http://dx.doi.org/10.1103/PhysRevD.23.347}{\emph{Phys.Rev.} {\bf D23}
  (1981) 347--356}.

\bibitem{Linde:1981mu}
A.~D. Linde, \emph{{A New Inflationary Universe Scenario: A Possible Solution
  of the Horizon, Flatness, Homogeneity, Isotropy and Primordial Monopole
  Problems}},
  \href{http://dx.doi.org/10.1016/0370-2693(82)91219-9}{\emph{Phys.Lett.} {\bf
  B108} (1982) 389--393}.

\bibitem{Albrecht:1982wi}
A.~Albrecht and P.~J. Steinhardt, \emph{{Cosmology for Grand Unified Theories
  with Radiatively Induced Symmetry Breaking}},
  \href{http://dx.doi.org/10.1103/PhysRevLett.48.1220}{\emph{Phys.Rev.Lett.}
  {\bf 48} (1982) 1220--1223}.

\bibitem{Linde:1983gd}
A.~D. Linde, \emph{{Chaotic Inflation}},
  \href{http://dx.doi.org/10.1016/0370-2693(83)90837-7}{\emph{Phys.Lett.} {\bf
  B129} (1983) 177--181}.

\bibitem{Mukhanov:1981xt}
V.~F. Mukhanov and G.~Chibisov, \emph{{Quantum Fluctuation and Nonsingular
  Universe.}}, {\emph{JETP Lett.} {\bf 33} (1981) 532--535}.

\bibitem{Hawking:1982cz}
S.~Hawking, \emph{{The Development of Irregularities in a Single Bubble
  Inflationary Universe}},
  \href{http://dx.doi.org/10.1016/0370-2693(82)90373-2}{\emph{Phys.Lett.} {\bf
  B115} (1982) 295}.

\bibitem{Starobinsky:1982ee}
A.~A. Starobinsky, \emph{{Dynamics of Phase Transition in the New Inflationary
  Universe Scenario and Generation of Perturbations}},
  \href{http://dx.doi.org/10.1016/0370-2693(82)90541-X}{\emph{Phys.Lett.} {\bf
  B117} (1982) 175--178}.

\bibitem{Guth:1982ec}
A.~H. Guth and S.~Pi, \emph{{Fluctuations in the New Inflationary Universe}},
  \href{http://dx.doi.org/10.1103/PhysRevLett.49.1110}{\emph{Phys.Rev.Lett.}
  {\bf 49} (1982) 1110--1113}.

\bibitem{Bardeen:1983qw}
J.~M. Bardeen, P.~J. Steinhardt and M.~S. Turner, \emph{{Spontaneous Creation
  of Almost Scale - Free Density Perturbations in an Inflationary Universe}},
  \href{http://dx.doi.org/10.1103/PhysRevD.28.679}{\emph{Phys.Rev.} {\bf D28}
  (1983) 679}.

\bibitem{Starobinsky:1979ty}
A.~A. Starobinsky, \emph{{Spectrum of relict gravitational radiation and the
  early state of the universe}}, {\emph{JETP Lett.} {\bf 30} (1979) 682--685}.

\bibitem{Polarski:1995jg}
D.~Polarski and A.~A. Starobinsky, \emph{{Semiclassicality and decoherence of
  cosmological perturbations}},
  \href{http://dx.doi.org/10.1088/0264-9381/13/3/006}{\emph{Class.Quant.Grav.}
  {\bf 13} (1996) 377--392}, [\href{http://arxiv.org/abs/gr-qc/9504030}{{\tt
  gr-qc/9504030}}].

\bibitem{Lesgourgues:1996jc}
J.~Lesgourgues, D.~Polarski and A.~A. Starobinsky, \emph{{Quantum to classical
  transition of cosmological perturbations for nonvacuum initial states}},
  \href{http://dx.doi.org/10.1016/S0550-3213(97)00224-1}{\emph{Nucl.Phys.} {\bf
  B497} (1997) 479--510}, [\href{http://arxiv.org/abs/gr-qc/9611019}{{\tt
  gr-qc/9611019}}].

\bibitem{Kiefer:2008ku}
C.~Kiefer and D.~Polarski, \emph{{Why do cosmological perturbations look
  classical to us?}},
  \href{http://dx.doi.org/10.1166/asl.2009.1023}{\emph{Adv.Sci.Lett.} {\bf 2}
  (2009) 164--173}, [\href{http://arxiv.org/abs/0810.0087}{{\tt 0810.0087}}].

\bibitem{Martin:2012pea}
J.~Martin, V.~Vennin and P.~Peter, \emph{{Cosmological Inflation and the
  Quantum Measurement Problem}},
  \href{http://dx.doi.org/10.1103/PhysRevD.86.103524}{\emph{Phys.Rev.} {\bf
  D86} (2012) 103524}, [\href{http://arxiv.org/abs/1207.2086}{{\tt
  1207.2086}}].

\bibitem{Burgess:2014eoa}
C.~P. Burgess, R.~Holman, G.~Tasinato and M.~Williams, \emph{{EFT Beyond the
  Horizon: Stochastic Inflation and How Primordial Quantum Fluctuations Go
  Classical}}, \href{http://dx.doi.org/10.1007/JHEP03(2015)090}{\emph{JHEP}
  {\bf 03} (2015) 090}, [\href{http://arxiv.org/abs/1408.5002}{{\tt
  1408.5002}}].

\bibitem{Martin:2015qta}
J.~Martin and V.~Vennin, \emph{{Quantum Discord of Cosmic Inflation: Can we
  Show that CMB Anisotropies are of Quantum-Mechanical Origin?}},
  \href{http://dx.doi.org/10.1103/PhysRevD.93.023505}{\emph{Phys. Rev.} {\bf
  D93} (2016) 023505}, [\href{http://arxiv.org/abs/1510.04038}{{\tt
  1510.04038}}].

\bibitem{Starobinsky:1986fx}
A.~A. Starobinsky, \emph{{Stochastic de Sitter (inflationary) stage in the
  early universe}},
  \href{http://dx.doi.org/10.1007/3-540-16452-9_6}{\emph{Lect.Notes Phys.} {\bf
  246} (1986) 107--126}.

\bibitem{Nambu:1987ef}
Y.~Nambu and M.~Sasaki, \emph{{Stochastic Stage of an Inflationary Universe
  Model}},
  \href{http://dx.doi.org/10.1016/0370-2693(88)90974-4}{\emph{Phys.Lett.} {\bf
  B205} (1988) 441}.

\bibitem{Nambu:1988je}
Y.~Nambu and M.~Sasaki, \emph{{Stochastic Approach to Chaotic Inflation and the
  Distribution of Universes}},
  \href{http://dx.doi.org/10.1016/0370-2693(89)90385-7}{\emph{Phys.Lett.} {\bf
  B219} (1989) 240}.

\bibitem{Kandrup:1988sc}
H.~E. Kandrup, \emph{{Stochastic inflation as a time dependent random walk}},
  \href{http://dx.doi.org/10.1103/PhysRevD.39.2245}{\emph{Phys.Rev.} {\bf D39}
  (1989) 2245}.

\bibitem{Nakao:1988yi}
K.-i. Nakao, Y.~Nambu and M.~Sasaki, \emph{{Stochastic Dynamics of New
  Inflation}},
  \href{http://dx.doi.org/10.1143/PTP.80.1041}{\emph{Prog.Theor.Phys.} {\bf 80}
  (1988) 1041}.

\bibitem{Nambu:1989uf}
Y.~Nambu, \emph{{Stochastic Dynamics of an Inflationary Model and Initial
  Distribution of Universes}},
  \href{http://dx.doi.org/10.1143/PTP.81.1037}{\emph{Prog.Theor.Phys.} {\bf 81}
  (1989) 1037}.

\bibitem{Mollerach:1990zf}
S.~Mollerach, S.~Matarrese, A.~Ortolan and F.~Lucchin, \emph{{Stochastic
  inflation in a simple two field model}},
  \href{http://dx.doi.org/10.1103/PhysRevD.44.1670}{\emph{Phys.Rev.} {\bf D44}
  (1991) 1670--1679}.

\bibitem{Linde:1993xx}
A.~D. Linde, D.~A. Linde and A.~Mezhlumian, \emph{{From the Big Bang theory to
  the theory of a stationary universe}},
  \href{http://dx.doi.org/10.1103/PhysRevD.49.1783}{\emph{Phys.Rev.} {\bf D49}
  (1994) 1783--1826}, [\href{http://arxiv.org/abs/gr-qc/9306035}{{\tt
  gr-qc/9306035}}].

\bibitem{Starobinsky:1994bd}
A.~A. Starobinsky and J.~Yokoyama, \emph{{Equilibrium state of a
  selfinteracting scalar field in the De Sitter background}},
  \href{http://dx.doi.org/10.1103/PhysRevD.50.6357}{\emph{Phys.Rev.} {\bf D50}
  (1994) 6357--6368}, [\href{http://arxiv.org/abs/astro-ph/9407016}{{\tt
  astro-ph/9407016}}].

\bibitem{GarciaBellido:1993wn}
J.~Garcia-Bellido, A.~D. Linde and D.~A. Linde, \emph{{Fluctuations of the
  gravitational constant in the inflationary Brans-Dicke cosmology}},
  \href{http://dx.doi.org/10.1103/PhysRevD.50.730}{\emph{Phys. Rev.} {\bf D50}
  (1994) 730--750}, [\href{http://arxiv.org/abs/astro-ph/9312039}{{\tt
  astro-ph/9312039}}].

\bibitem{Kahya:2006hc}
E.~O. Kahya and V.~K. Onemli, \emph{{Quantum Stability of a $w < -1$ Phase of
  Cosmic Acceleration}},
  \href{http://dx.doi.org/10.1103/PhysRevD.76.043512}{\emph{Phys. Rev.} {\bf
  D76} (2007) 043512}, [\href{http://arxiv.org/abs/gr-qc/0612026}{{\tt
  gr-qc/0612026}}].

\bibitem{Finelli:2008zg}
F.~Finelli, G.~Marozzi, A.~Starobinsky, G.~Vacca and G.~Venturi,
  \emph{{Generation of fluctuations during inflation: Comparison of stochastic
  and field-theoretic approaches}},
  \href{http://dx.doi.org/10.1103/PhysRevD.79.044007}{\emph{Phys.Rev.} {\bf
  D79} (2009) 044007}, [\href{http://arxiv.org/abs/0808.1786}{{\tt
  0808.1786}}].

\bibitem{Finelli:2010sh}
F.~Finelli, G.~Marozzi, A.~Starobinsky, G.~Vacca and G.~Venturi,
  \emph{{Stochastic growth of quantum fluctuations during slow-roll
  inflation}},
  \href{http://dx.doi.org/10.1103/PhysRevD.82.064020}{\emph{Phys.Rev.} {\bf
  D82} (2010) 064020}, [\href{http://arxiv.org/abs/1003.1327}{{\tt
  1003.1327}}].

\bibitem{Garbrecht:2013coa}
B.~Garbrecht, G.~Rigopoulos and Y.~Zhu, \emph{{Infrared Correlations in de
  Sitter Space: Field Theoretic vs. Stochastic Approach}}, {\emph{Phys.Rev.}
  {\bf D89} (2014) 063506}, [\href{http://arxiv.org/abs/1310.0367}{{\tt
  1310.0367}}].

\bibitem{Garbrecht:2014dca}
B.~Garbrecht, F.~Gautier, G.~Rigopoulos and Y.~Zhu, \emph{{Feynman Diagrams for
  Stochastic Inflation and Quantum Field Theory in de Sitter Space}},
  \href{http://dx.doi.org/10.1103/PhysRevD.91.063520}{\emph{Phys. Rev.} {\bf
  D91} (2015) 063520}, [\href{http://arxiv.org/abs/1412.4893}{{\tt
  1412.4893}}].

\bibitem{Levasseur:2014ska}
L.~P. Levasseur and E.~McDonough, \emph{{Backreaction and Stochastic Effects in
  Single Field Inflation}},
  \href{http://dx.doi.org/10.1103/PhysRevD.91.063513}{\emph{Phys. Rev.} {\bf
  D91} (2015) 063513}, [\href{http://arxiv.org/abs/1409.7399}{{\tt
  1409.7399}}].

\bibitem{Onemli:2015pma}
V.~K. Onemli, \emph{{Vacuum Fluctuations of a Scalar Field during Inflation:
  Quantum versus Stochastic Analysis}},
  \href{http://dx.doi.org/10.1103/PhysRevD.91.103537}{\emph{Phys. Rev.} {\bf
  D91} (2015) 103537}, [\href{http://arxiv.org/abs/1501.05852}{{\tt
  1501.05852}}].

\bibitem{Prokopec:2015owa}
T.~Prokopec, \emph{{Late time solution for interacting scalar in accelerating
  spaces}}, \href{http://dx.doi.org/10.1088/1475-7516/2015/11/016}{\emph{JCAP}
  {\bf 1511} (2015) 016}, [\href{http://arxiv.org/abs/1508.07874}{{\tt
  1508.07874}}].

\bibitem{Boyanovsky:2015jen}
D.~Boyanovsky, \emph{{Effective field theory during inflation. II. Stochastic
  dynamics and power spectrum suppression}},
  \href{http://dx.doi.org/10.1103/PhysRevD.93.043501}{\emph{Phys. Rev.} {\bf
  D93} (2016) 043501}, [\href{http://arxiv.org/abs/1511.06649}{{\tt
  1511.06649}}].

\bibitem{Boyanovsky:2015xoa}
D.~Boyanovsky, \emph{{Effective Field Theory out of Equilibrium: Brownian
  quantum fields}},
  \href{http://dx.doi.org/10.1088/1367-2630/17/6/063017}{\emph{New J. Phys.}
  {\bf 17} (2015) 063017}, [\href{http://arxiv.org/abs/1503.00156}{{\tt
  1503.00156}}].

\bibitem{Boyanovsky:2015tba}
D.~Boyanovsky, \emph{{Effective field theory during inflation: Reduced density
  matrix and its quantum master equation}},
  \href{http://dx.doi.org/10.1103/PhysRevD.92.023527}{\emph{Phys. Rev.} {\bf
  D92} (2015) 023527}, [\href{http://arxiv.org/abs/1506.07395}{{\tt
  1506.07395}}].

\bibitem{Burgess:2015ajz}
C.~P. Burgess, R.~Holman and G.~Tasinato, \emph{{Open EFTs, IR effects \&
  late-time resummations: systematic corrections in stochastic inflation}},
  \href{http://dx.doi.org/10.1007/JHEP01(2016)153}{\emph{JHEP} {\bf 01} (2016)
  153}, [\href{http://arxiv.org/abs/1512.00169}{{\tt 1512.00169}}].

\bibitem{Prokopec:2007ak}
T.~Prokopec, N.~Tsamis and R.~Woodard, \emph{{Stochastic Inflationary Scalar
  Electrodynamics}},
  \href{http://dx.doi.org/10.1016/j.aop.2007.08.008}{\emph{Annals Phys.} {\bf
  323} (2008) 1324--1360}, [\href{http://arxiv.org/abs/0707.0847}{{\tt
  0707.0847}}].

\bibitem{Prokopec:2008gw}
T.~Prokopec, N.~C. Tsamis and R.~P. Woodard, \emph{{Two loop stress-energy
  tensor for inflationary scalar electrodynamics}},
  \href{http://dx.doi.org/10.1103/PhysRevD.78.043523}{\emph{Phys.Rev.} {\bf
  D78} (2008) 043523}, [\href{http://arxiv.org/abs/0802.3673}{{\tt
  0802.3673}}].

\bibitem{Tsamis:2005hd}
N.~Tsamis and R.~Woodard, \emph{{Stochastic quantum gravitational inflation}},
  \href{http://dx.doi.org/10.1016/j.nuclphysb.2005.06.031}{\emph{Nucl.Phys.}
  {\bf B724} (2005) 295--328}, [\href{http://arxiv.org/abs/gr-qc/0505115}{{\tt
  gr-qc/0505115}}].

\bibitem{Starobinsky:1986fxa}
A.~A. Starobinsky, \emph{{Multicomponent de Sitter (Inflationary) Stages and
  the Generation of Perturbations}}, {\emph{JETP Lett.} {\bf 42} (1985)
  152--155}.

\bibitem{Salopek:1990jq}
D.~S. Salopek and J.~R. Bond, \emph{{Nonlinear evolution of long wavelength
  metric fluctuations in inflationary models}},
  \href{http://dx.doi.org/10.1103/PhysRevD.42.3936}{\emph{Phys. Rev.} {\bf D42}
  (1990) 3936--3962}.

\bibitem{Sasaki:1995aw}
M.~Sasaki and E.~D. Stewart, \emph{{A General analytic formula for the spectral
  index of the density perturbations produced during inflation}},
  \href{http://dx.doi.org/10.1143/PTP.95.71}{\emph{Prog.Theor.Phys.} {\bf 95}
  (1996) 71--78}, [\href{http://arxiv.org/abs/astro-ph/9507001}{{\tt
  astro-ph/9507001}}].

\bibitem{Sasaki:1998ug}
M.~Sasaki and T.~Tanaka, \emph{{Superhorizon scale dynamics of multiscalar
  inflation}},
  \href{http://dx.doi.org/10.1143/PTP.99.763}{\emph{Prog.Theor.Phys.} {\bf 99}
  (1998) 763--782}, [\href{http://arxiv.org/abs/gr-qc/9801017}{{\tt
  gr-qc/9801017}}].

\bibitem{Wands:2000dp}
D.~Wands, K.~A. Malik, D.~H. Lyth and A.~R. Liddle, \emph{{A New approach to
  the evolution of cosmological perturbations on large scales}},
  \href{http://dx.doi.org/10.1103/PhysRevD.62.043527}{\emph{Phys.Rev.} {\bf
  D62} (2000) 043527}, [\href{http://arxiv.org/abs/astro-ph/0003278}{{\tt
  astro-ph/0003278}}].

\bibitem{Lyth:2004gb}
D.~H. Lyth, K.~A. Malik and M.~Sasaki, \emph{{A General proof of the
  conservation of the curvature perturbation}},
  \href{http://dx.doi.org/10.1088/1475-7516/2005/05/004}{\emph{JCAP} {\bf 0505}
  (2005) 004}, [\href{http://arxiv.org/abs/astro-ph/0411220}{{\tt
  astro-ph/0411220}}].

\bibitem{Lyth:2005fi}
D.~H. Lyth and Y.~Rodriguez, \emph{{The Inflationary prediction for primordial
  non-Gaussianity}},
  \href{http://dx.doi.org/10.1103/PhysRevLett.95.121302}{\emph{Phys.Rev.Lett.}
  {\bf 95} (2005) 121302}, [\href{http://arxiv.org/abs/astro-ph/0504045}{{\tt
  astro-ph/0504045}}].

\bibitem{Enqvist:2008kt}
K.~Enqvist, S.~Nurmi, D.~Podolsky and G.~Rigopoulos, \emph{{On the divergences
  of inflationary superhorizon perturbations}},
  \href{http://dx.doi.org/10.1088/1475-7516/2008/04/025}{\emph{JCAP} {\bf 0804}
  (2008) 025}, [\href{http://arxiv.org/abs/0802.0395}{{\tt 0802.0395}}].

\bibitem{Fujita:2013cna}
T.~Fujita, M.~Kawasaki, Y.~Tada and T.~Takesako, \emph{{A new algorithm for
  calculating the curvature perturbations in stochastic inflation}},
  \href{http://dx.doi.org/10.1088/1475-7516/2013/12/036}{\emph{JCAP} {\bf 1312}
  (2013) 036}, [\href{http://arxiv.org/abs/1308.4754}{{\tt 1308.4754}}].

\bibitem{Fujita:2014tja}
T.~Fujita, M.~Kawasaki and Y.~Tada, \emph{{Non-perturbative approach for
  curvature perturbations in stochastic-$\delta N$ formalism}},
  \href{http://arxiv.org/abs/1405.2187}{{\tt 1405.2187}}.

\bibitem{Vennin:2015hra}
V.~Vennin and A.~A. Starobinsky, \emph{{Correlation Functions in Stochastic
  Inflation}},
  \href{http://dx.doi.org/10.1140/epjc/s10052-015-3643-y}{\emph{Eur. Phys. J.}
  {\bf C75} (2015) 413}, [\href{http://arxiv.org/abs/1506.04732}{{\tt
  1506.04732}}].

\bibitem{Kawasaki:2015ppx}
M.~Kawasaki and Y.~Tada, \emph{{Can massive primordial black holes be produced
  in mild waterfall hybrid inflation?}},
  \href{http://arxiv.org/abs/1512.03515}{{\tt 1512.03515}}.

\bibitem{Clesse:2009ur}
S.~Clesse, C.~Ringeval and J.~Rocher, \emph{{Fractal initial conditions and
  natural parameter values in hybrid inflation}},
  \href{http://dx.doi.org/10.1103/PhysRevD.80.123534}{\emph{Phys. Rev.} {\bf
  D80} (2009) 123534}, [\href{http://arxiv.org/abs/0909.0402}{{\tt
  0909.0402}}].

\bibitem{Martin:2011ib}
J.~Martin and V.~Vennin, \emph{{Stochastic Effects in Hybrid Inflation}},
  \href{http://dx.doi.org/10.1103/PhysRevD.85.043525}{\emph{Phys. Rev.} {\bf
  D85} (2012) 043525}, [\href{http://arxiv.org/abs/1110.2070}{{\tt
  1110.2070}}].

\bibitem{Levasseur:2013tja}
L.~Perreault~Levasseur, V.~Vennin and R.~Brandenberger, \emph{{Recursive
  Stochastic Effects in Valley Hybrid Inflation}},
  \href{http://dx.doi.org/10.1103/PhysRevD.88.083538}{\emph{Phys. Rev.} {\bf
  D88} (2013) 083538}, [\href{http://arxiv.org/abs/1307.2575}{{\tt
  1307.2575}}].

\bibitem{Stratonovich:1966}
R.~L. Stratonovich, \emph{{A New Representation for Stochastic Integrals and
  Equations}}, \href{http://dx.doi.org/10.1137/0304028}{\emph{SIAM Journal on
  Control} {\bf 4} (1966) 362--371}.

\bibitem{Winitzki:1995pg}
S.~Winitzki and A.~Vilenkin, \emph{{Uncertainties of predictions in models of
  eternal inflation}},
  \href{http://dx.doi.org/10.1103/PhysRevD.53.4298}{\emph{Phys.Rev.} {\bf D53}
  (1996) 4298--4310}, [\href{http://arxiv.org/abs/gr-qc/9510054}{{\tt
  gr-qc/9510054}}].

\bibitem{Vilenkin:1999kd}
A.~Vilenkin, \emph{{On the factor ordering problem in stochastic inflation}},
  \href{http://dx.doi.org/10.1103/PhysRevD.59.123506}{\emph{Phys.Rev.} {\bf
  D59} (1999) 123506}, [\href{http://arxiv.org/abs/gr-qc/9902007}{{\tt
  gr-qc/9902007}}].

\bibitem{Linde:1993cn}
A.~D. Linde, \emph{{Hybrid inflation}},
  \href{http://dx.doi.org/10.1103/PhysRevD.49.748}{\emph{Phys. Rev.} {\bf D49}
  (1994) 748--754}, [\href{http://arxiv.org/abs/astro-ph/9307002}{{\tt
  astro-ph/9307002}}].

\bibitem{Copeland:1994vg}
E.~J. Copeland, A.~R. Liddle, D.~H. Lyth, E.~D. Stewart and D.~Wands,
  \emph{{False vacuum inflation with Einstein gravity}},
  \href{http://dx.doi.org/10.1103/PhysRevD.49.6410}{\emph{Phys. Rev.} {\bf D49}
  (1994) 6410--6433}, [\href{http://arxiv.org/abs/astro-ph/9401011}{{\tt
  astro-ph/9401011}}].

\bibitem{Bernardeau:2002jf}
F.~Bernardeau and J.-P. Uzan, \emph{{Inflationary models inducing non-Gaussian
  metric fluctuations}},
  \href{http://dx.doi.org/10.1103/PhysRevD.67.121301}{\emph{Phys. Rev.} {\bf
  D67} (2003) 121301}, [\href{http://arxiv.org/abs/astro-ph/0209330}{{\tt
  astro-ph/0209330}}].

\bibitem{Bernardeau:2004zz}
F.~Bernardeau, L.~Kofman and J.-P. Uzan, \emph{{Modulated fluctuations from
  hybrid inflation}},
  \href{http://dx.doi.org/10.1103/PhysRevD.70.083004}{\emph{Phys. Rev.} {\bf
  D70} (2004) 083004}, [\href{http://arxiv.org/abs/astro-ph/0403315}{{\tt
  astro-ph/0403315}}].

\bibitem{Lyth:2005qk}
D.~H. Lyth, \emph{{Generating the curvature perturbation at the end of
  inflation}},
  \href{http://dx.doi.org/10.1088/1475-7516/2005/11/006}{\emph{JCAP} {\bf 0511}
  (2005) 006}, [\href{http://arxiv.org/abs/astro-ph/0510443}{{\tt
  astro-ph/0510443}}].

\bibitem{Alabidi:2006wa}
L.~Alabidi and D.~Lyth, \emph{{Curvature perturbation from symmetry breaking
  the end of inflation}},
  \href{http://dx.doi.org/10.1088/1475-7516/2006/08/006}{\emph{JCAP} {\bf 0608}
  (2006) 006}, [\href{http://arxiv.org/abs/astro-ph/0604569}{{\tt
  astro-ph/0604569}}].

\bibitem{Bachelier:1900}
L.~Bachelier, \emph{{Theorie de la speculation}}.
\newblock {Gauthier-Villars}, 1900.

\bibitem{Gihman:1972}
I.~Gihman and A.~Skorohod, \emph{{Stochastic Differential Equations}}.
\newblock {Springer Verlag, Berlin Heidelberg New York}, 1972, p.108.

\bibitem{Gordon:2000hv}
C.~Gordon, D.~Wands, B.~A. Bassett and R.~Maartens, \emph{{Adiabatic and
  entropy perturbations from inflation}},
  \href{http://dx.doi.org/10.1103/PhysRevD.63.023506}{\emph{Phys. Rev.} {\bf
  D63} (2001) 023506}, [\href{http://arxiv.org/abs/astro-ph/0009131}{{\tt
  astro-ph/0009131}}].

\bibitem{Malik:1998gy}
K.~A. Malik and D.~Wands, \emph{{Dynamics of assisted inflation}},
  \href{http://dx.doi.org/10.1103/PhysRevD.59.123501}{\emph{Phys. Rev.} {\bf
  D59} (1999) 123501}, [\href{http://arxiv.org/abs/astro-ph/9812204}{{\tt
  astro-ph/9812204}}].

\bibitem{Saffin:2012et}
P.~M. Saffin, \emph{{The covariance of multi-field perturbations, pseudo-susy
  and fNL}}, \href{http://dx.doi.org/10.1088/1475-7516/2012/09/002}{\emph{JCAP}
  {\bf 1209} (2012) 002}, [\href{http://arxiv.org/abs/1203.0397}{{\tt
  1203.0397}}].

\bibitem{Steinhardt:1982kg}
P.~J. Steinhardt, \emph{{NATURAL INFLATION}},  in \emph{{In *Cambridge 1982,
  Proceedings, The Very Early Universe*, 251-266 and Pennsylvania Univ.
  Philadelphia - UPR-0198T (82,REC.OCT.) 16 P. (211400) (SEE CONFERENCE
  INDEX)}}, 1982.

\bibitem{Vilenkin:1983xq}
A.~Vilenkin, \emph{{The Birth of Inflationary Universes}},
  \href{http://dx.doi.org/10.1103/PhysRevD.27.2848}{\emph{Phys. Rev.} {\bf D27}
  (1983) 2848}.

\bibitem{Guth:1985ya}
A.~H. Guth and S.-Y. Pi, \emph{{The Quantum Mechanics of the Scalar Field in
  the New Inflationary Universe}},
  \href{http://dx.doi.org/10.1103/PhysRevD.32.1899}{\emph{Phys. Rev.} {\bf D32}
  (1985) 1899--1920}.

\bibitem{Linde:1986fc}
A.~D. Linde, \emph{{ETERNAL CHAOTIC INFLATION}},
  \href{http://dx.doi.org/10.1142/S0217732386000129}{\emph{Mod. Phys. Lett.}
  {\bf A1} (1986) 81}.

\bibitem{Barenboim:2016mmw}
G.~Barenboim, W.~H. Kinney and W.-I. Park, \emph{{Eternal Hilltop Inflation}},
  \href{http://arxiv.org/abs/1601.08140}{{\tt 1601.08140}}.

\bibitem{Polya:1921}
G.~P\'olya, \emph{{\"Uber eine aufgabe betreffend die irrfahrt im
  strassennetz}}, {\emph{Math. Ann.} {\bf 84} (1921) 149–160}.

\bibitem{Polarski:1992dq}
D.~Polarski and A.~A. Starobinsky, \emph{{Spectra of perturbations produced by
  double inflation with an intermediate matter dominated stage}},
  \href{http://dx.doi.org/10.1016/0550-3213(92)90062-G}{\emph{Nucl. Phys.} {\bf
  B385} (1992) 623--650}.

\bibitem{Ringeval:2010hf}
C.~Ringeval, T.~Suyama, T.~Takahashi, M.~Yamaguchi and S.~Yokoyama, \emph{{Dark
  energy from primordial inflationary quantum fluctuations}},
  \href{http://dx.doi.org/10.1103/PhysRevLett.105.121301}{\emph{Phys. Rev.
  Lett.} {\bf 105} (2010) 121301}, [\href{http://arxiv.org/abs/1006.0368}{{\tt
  1006.0368}}].

\bibitem{Enqvist:2012xn}
K.~Enqvist, R.~N. Lerner, O.~Taanila and A.~Tranberg, \emph{{Spectator field
  dynamics in de Sitter and curvaton initial conditions}},
  \href{http://dx.doi.org/10.1088/1475-7516/2012/10/052}{\emph{JCAP} {\bf 1210}
  (2012) 052}, [\href{http://arxiv.org/abs/1205.5446}{{\tt 1205.5446}}].

\bibitem{Vennin:2016wnk}
V.~Vennin, H.~Assadullahi, H.~Firouzjahi, M.~Noorbala and D.~Wands,
  \emph{{Critical Number of Fields in Stochastic Inflation}},
  \href{http://arxiv.org/abs/1604.06017}{{\tt 1604.06017}}.

\bibitem{Ade:2015xua}
{\scshape Planck} collaboration, P.~A.~R. Ade et~al., \emph{{Planck 2015
  results. XIII. Cosmological parameters}},
  \href{http://arxiv.org/abs/1502.01589}{{\tt 1502.01589}}.

\bibitem{Broy:2014sia}
B.~J. Broy, D.~Roest and A.~Westphal, \emph{{Power Spectrum of Inflationary
  Attractors}}, \href{http://dx.doi.org/10.1103/PhysRevD.91.023514}{\emph{Phys.
  Rev.} {\bf D91} (2015) 023514}, [\href{http://arxiv.org/abs/1408.5904}{{\tt
  1408.5904}}].

\bibitem{Coone:2015fha}
D.~Coone, D.~Roest and V.~Vennin, \emph{{The Hubble Flow of Plateau
  Inflation}},
  \href{http://dx.doi.org/10.1088/1475-7516/2015/11/010}{\emph{JCAP} {\bf 1511}
  (2015) 010}, [\href{http://arxiv.org/abs/1507.00096}{{\tt 1507.00096}}].

\bibitem{Dvali:2003ar}
G.~Dvali, A.~Gruzinov and M.~Zaldarriaga, \emph{{Cosmological perturbations
  from inhomogeneous reheating, freezeout, and mass domination}},
  \href{http://dx.doi.org/10.1103/PhysRevD.69.083505}{\emph{Phys. Rev.} {\bf
  D69} (2004) 083505}, [\href{http://arxiv.org/abs/astro-ph/0305548}{{\tt
  astro-ph/0305548}}].

\bibitem{Abbott:1984fp}
L.~F. Abbott and M.~B. Wise, \emph{{Constraints on Generalized Inflationary
  Cosmologies}},
  \href{http://dx.doi.org/10.1016/0550-3213(84)90329-8}{\emph{Nucl. Phys.} {\bf
  B244} (1984) 541--548}.

\bibitem{Vennin:2015egh}
V.~Vennin, K.~Koyama and D.~Wands, \emph{{Inflation with an extra light scalar
  field after Planck}},  \href{http://arxiv.org/abs/1512.03403}{{\tt
  1512.03403}}.

\bibitem{ito1944}
K.~It\^o, \emph{Stochastic integral},
  \href{http://dx.doi.org/10.3792/pia/1195572786}{\emph{Proceedings of the
  Imperial Academy} {\bf 20} (1944) 519--524}.

\bibitem{Risken:1984book}
H.~Risken, \emph{{The Fokker-Planck Equation}}, vol.~18.
\newblock Spinger Series in Synergetics, 1984.

\end{thebibliography}\endgroup
\end{document}